\documentclass[sigconf]{acmart}

\usepackage{subfig}
\usepackage{multirow}
\usepackage{multicol}
\usepackage{enumitem}
\usepackage[normalem]{ulem}
\usepackage[listings]{tcolorbox}
\tcbuselibrary{listings,theorems}
\usepackage{colortbl}
\usepackage{xcolor}

\newcommand{\colorgray}{\cellcolor[rgb]{0.906, 0.902, 0.902}}
\newcommand{\thanksintern}{\authornote{Work done during the internship at Microsoft.}}
\AtBeginDocument{%
  \providecommand\BibTeX{{%
    \normalfont B\kern-0.5em{\scshape i\kern-0.25em b}\kern-0.8em\TeX}}}

\setcopyright{acmlicensed}
\copyrightyear{2024}
\acmYear{2024}
\acmDOI{10.1145/3643916.3644403}

\acmConference[ICPC '24]{32nd IEEE/ACM International Conference on Program Comprehension}{April 15--16, 2024}{Lisbon, Portugal}
\acmBooktitle{32nd IEEE/ACM International Conference on Program Comprehension (ICPC '24), April 15--16, 2024, Lisbon, Portugal}
\acmISBN{979-8-4007-0586-1/24/04}

\newcommand{\find}[1]{
\begin{tcolorbox}[leftrule=1mm,rightrule=1mm,toprule=0mm,bottomrule=0mm,left=1pt,right=1pt,top=0.5pt,bottom=0.5pt]
\em #1
\end{tcolorbox}
}
\definecolor{purple1}{HTML}{8d3a94}

\newtcbtheorem[]{exmp}{Prompt}%
{colback=purple1!5,colframe=purple1!80,fonttitle=\bfseries, left=2pt, right=2pt,top=2pt,bottom=2pt}{exmp}

\begin{document}

\title[Compositional API Recommendation for Library-Oriented Code Generation]{
Compositional API Recommendation for\\Library-Oriented Code Generation}


\author{Zexiong Ma}
\thanksintern
\affiliation{%
  \institution{School of Computer Science, Peking University\\ Key Laboratory of High Conidence Software Technologies, Ministry of Education}
  \city{Beijing}
  \country{China}}
\email{mazexiong@stu.pku.edu.cn}

\author{Shengnan An}
\authornotemark[1]
\affiliation{%
  \institution{Xi'an Jiaotong University}
  \city{Xi'an}
  \country{China}}
\email{an1006634493@stu.xjtu.edu.cn}

\author{Bing Xie}
\affiliation{%
  \institution{School of Computer Science, Peking University\\ Key Laboratory of High Conidence Software Technologies, Ministry of Education}
  \city{Beijing}
  \country{China}}
\email{xiebing@pku.edu.cn}

\author{Zeqi Lin}
\authornote{Corresponding author.}
\affiliation{%
  \institution{Microsoft Corporation}
  \city{Beijing}
  \country{China}}
\email{Zeqi.Lin@microsoft.com}

\renewcommand{\shortauthors}{Ma and An, et al.}

\begin{abstract}

Large language models (LLMs) have achieved exceptional performance in code generation. However, the performance remains unsatisfactory in generating library-oriented code, especially for the libraries not present in the training data of LLMs.
Previous work utilizes API recommendation technology to help LLMs use libraries: it retrieves APIs related to the user requirements, then leverages them as context to prompt LLMs.
However, developmental requirements can be coarse-grained, requiring  a combination of multiple fine-grained APIs. This granularity inconsistency makes API recommendation a challenging task.

To address this, we propose \textbf{CAPIR} (Compositional API Recommendation), which adopts a "divide-and-conquer" strategy  to recommend APIs for coarse-grained requirements.
Specifically, CAPIR employs an LLM-based \textit{Decomposer} to break down a coarse-grained task description into several detailed subtasks.
Then, CAPIR applies an embedding-based \textit{Retriever} to identify relevant APIs corresponding to each subtask. Moreover, CAPIR leverages an LLM-based \textit{Reranker} to filter out redundant APIs and provides the final recommendation.

To facilitate the evaluation of API recommendation methods on coarse-grained requirements, we present two challenging benchmarks, \textbf{RAPID} (Recommend APIs based on Documentation) and \textbf{LOCG} (Library-Oriented Code Generation).
Experimental results on these benchmarks, demonstrate the effectiveness of CAPIR in comparison to existing baselines. 
Specifically, on RAPID's Torchdata-AR dataset, compared to the state-of-the-art API recommendation approach, CAPIR improves recall@5 from 18.7\% to 43.2\% and precision@5 from 15.5\% to 37.1\%. On LOCG's Torchdata-Code dataset, compared to code generation without API recommendation, CAPIR improves pass@100 from 16.0\% to 28.0\%.

\end{abstract}






\maketitle

\section{Introduction}

Large language models (LLMs)~\cite{floridi2020gpt} have demonstrated impressive performance in code generation tasks~\cite{chen2022codet, zhang2023coder, li2022competition}. LLM-based programming assistance tools, like Copilot~\cite{Copilot, chen2021evaluating} and ChatGPT~\cite{chatgpt}, have been widely applied in the development process.
These tools can leverage the knowledge learning from the training data and generate executable code for developers.

Library-oriented code generation~\cite{zan2022cert, ase2023codegen4libs},  which involves generating code based on APIs within a specific library, plays a significant role in enhancing practical development efficiency, but existing LLM-based programming tools have unsatisfactory performance in this scenario. Specifically, the LLM-based programming tools are unable to generate code that import libraries not present or rarely present in the training data of LLM.
For example, the training data of ChatGPT is collected from internet before September 2021. This makes ChatGPT unable to generate code that import libraries released after that, while new libraries are continually being released in the open-source community. Moreover, some libraries are domain-specific or used in enterprise development,  which also limits their representation in the training data for LLMs. 
Some recent surveys~\cite{ciniselli2023source, ase2023codegen4libs} show that when using programming assistance tools, developers often expect the tool to generate code that invokes specific third-party libraries.
When developers want to use such low-resource libraries, they can only learn about the usage of the APIs through official documentation or community forums~\cite{barke2023grounded}.

\begin{figure}
    \centering
    \subfloat[Without API Recommendation]{
    \label{fig:motivation1} 
    \includegraphics[width=0.35\textwidth]{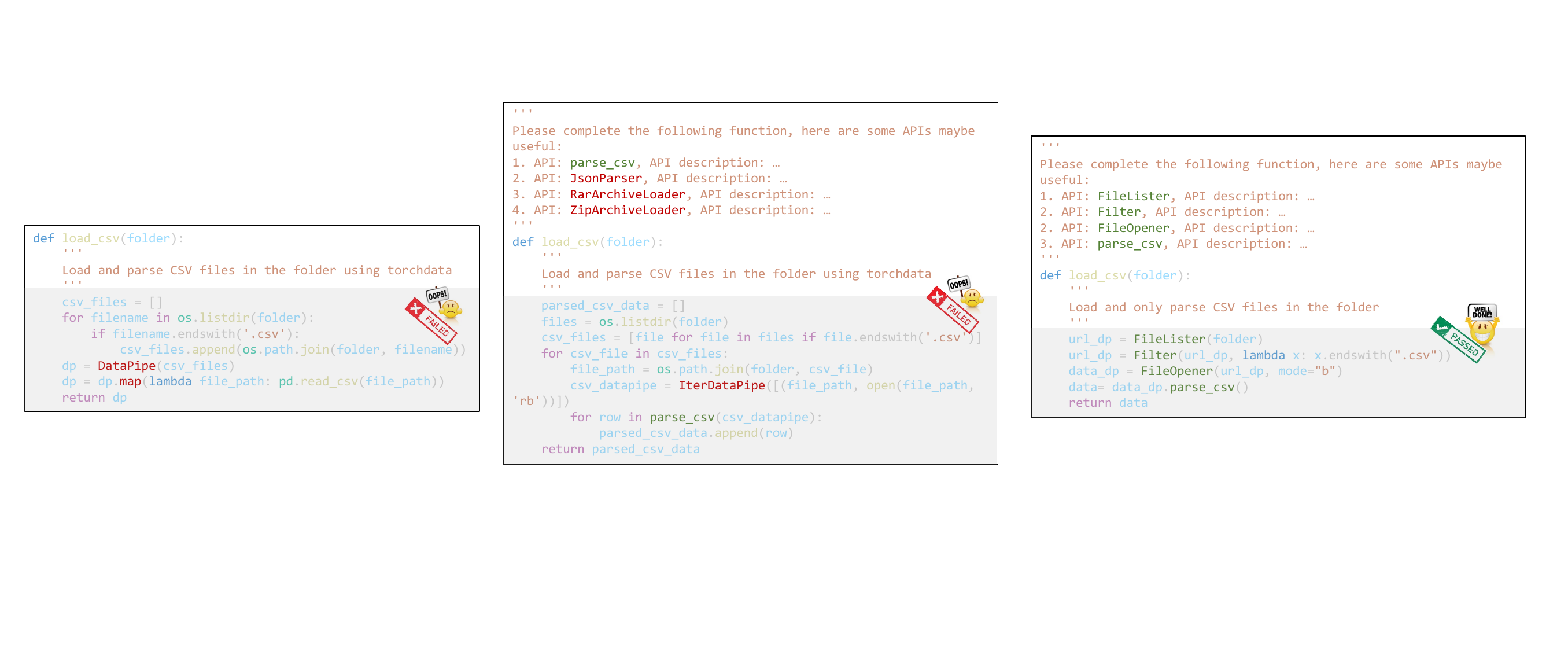}}\quad
    \subfloat[With API Recommendation directly based on requirements]{
    \label{fig:motivation2} 
    \includegraphics[width=0.35\textwidth]{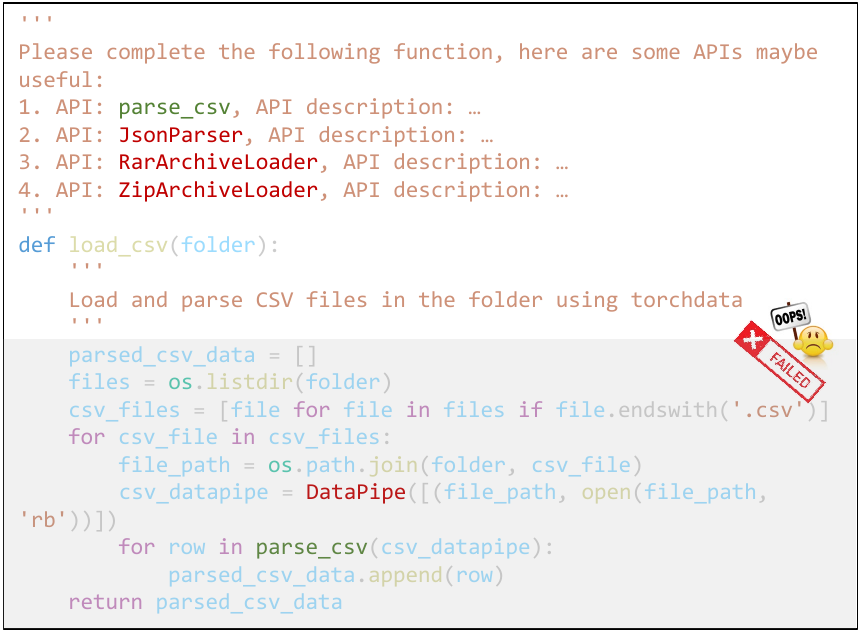}}\quad
  \subfloat[With Compositional API Recommendation (ours)]{
  \label{fig:motivation3} 
  \includegraphics[width=0.35\textwidth]{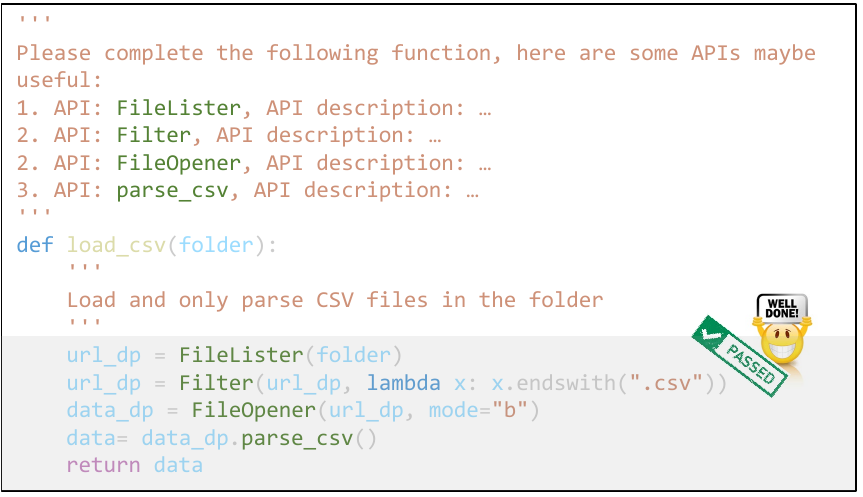}}
    \caption{An example of library-oriented code generation with LLMs.
    The \colorbox{gray!20}{gray parts} are the generated codes. 
    (a) Due to the lack of API information, LLM generates code that looks reasonable but cannot be executed correctly. (b) Directly using the requirement to retrieve APIs from document, LLM still generates the wrong code. (c) CAPIR Decomposes the task in to subtasks, retrieves APIs from document for each subtask, makes LLM generate the correct code.
    }
    \label{fig:noPrivateLibrary}
\end{figure}
Previous work~\cite{zhou2022docprompting, zan2023private} utilizes API recommendation to help LLMs invoke libraries, but existing approaches face two challenges for practical applications: 
coarse-grained requirements and lack relevant resources.
API recommendation has been an extensive research subject for many years~\cite{he2021pyart, peng2022revisiting, wei2022api, chan2012searching, irsan2023multi}. Its primary goal is to recommend APIs to users based on their developmental requirements. From a methodological perspective, these approaches can be broadly categorized into two types: \textit{learning-based methods} and \textit{retrieval-based methods}. However, when it comes to library-based code generation with LLMs, existing API recommendation methods encounter two challenges:
~\uline{ (1) The inconsistency in the granularity of requirements and APIs}. Some recent work ~\cite{zan2022language, zhou2022docprompting, zan2023private} has attempted to retrieve APIs from documents based on users' requirements and incorporate them into LLM prompts to assist in library-oriented code generation. Some recent surveys~\cite{vaithilingam2022expectation, barke2023grounded} show that when using programming assistance tools, users' requirements can be coarse-grained, needing a combination of multiple APIs. Directly retrieving from documentation may lead to unsatisfactory results. 
\uline{ (2) Insufficient training data}. Learning-based methods require expensive retraining for specific libraries. 
Some libraries lack relevant resources, making it challenging to obtain sufficient data for training.
For example, Torchdata
was released on Mar 11, 2022, which makes ChatGPT unable to generate related code. It is prohibitively expensive to retrain models specifically for Torchdata. Moreover, there is a scarcity of relevant trainable resources on the internet. As of April 10, 2023, there are only around 1k unduplicated related code files on GitHub. Additionally, there are only 3 posts tagged with Torchdata on StackOverflow
, which is insufficient for fine-tuning language models.
To help LLMs with library-oriented code generation, we need an API recommendation approach that does not require training data and can address coarse-grained tasks.

We present \textbf{CAPIR} (Compositional API Recommendation), a novel compositional API recommendation approach based on API documents. CAPIR utilizes the powerful reasoning capabilities of LLMs to decompose coarse-grained user requirements into fine-grained subtask sequence. These subtasks are then used to retrieve corresponding APIs from API documentation. 
While LLMs have demonstrated strong reasoning abilities~\cite{wei2022chain}, the key challenge lies in guiding the language model to decompose the requirement descriptions at the granularity of APIs in the target library. We use few-shot examples to demonstrate the decomposition granularity. 
CAPIR extracts code examples from development documents, and use $LLM_{s}$ (Summarizer) to generate summarizations~\cite{sun2023automatic} of these examples. The pairs of <summarization, API\_functions> serve as examples of <task, subtasks>. 
These <task, subtasks> examples are included in the prompt of $LLM_d$ (Decomposer) to guide its decomposition of requirements into API-level subtasks.
Then CAPIR uses an off-the-shelf embedding model as Retriever to retrieve APIs, and leverages  $LLM_r$ (Reranker) to rerank the API retrieval results. Specifically, we utilize ada-embedding-002 as Retriever, while employing gpt-3.5-turbo as the Summarizer, Decomposer, and Reranker.

In addition to the proposed CAPIR, this work introduces two benchmarks, \textbf{RAPID} and \textbf{LOCG}, to promote the evaluation of API recommendation and code generation for coarse-grained requirements. Specifically, RAPID encompasses four single-library API recommendation tasks and one multi-library task, and LOCG contains one single-library code generation task and one multi-library task. Experimental results on these two benchmarks demonstrate the effectiveness of CAPIR: In comparison to two strong baseline methods ADA-retrieve~\citep{ADA-embedding-002} and CLEAR~\citep{wei2022clear}, our CAPIR consistently achieves higher recall and precision in API recommendation tasks, and the recommended APIs lead to better pass@$k$ performance on library-oriented code generation tasks. Beyond these experimental results, we further validate the effectiveness of CAPIR for practical application, revealing that 
the recommended APIs from CAPIR can better assist human developers in accomplishing intricate coding tasks.

In summary, this paper makes the following main contributions:
\begin{itemize}
    \item
    We propose CAPIR, a novel compositional API recommendation approach, which only uses API documents and a few examples, allowing it to adapt to low-resource libraries.
    \item
    We present two challenging benchmarks, \textbf{RAPID} (Recommend APIs based on Documentation) and \textbf{LOCG} (Library-Oriented Code Generation). Experimental results on these benchmarks, demonstrate the effectiveness of CAPIR in comparison to existing baselines. 
    \item 
    We conduct a detailed qualitative analysis and perform experiments in real development scenarios, further validating that CAPIR can be applied for practical application.
\end{itemize}

\begin{figure*}
     \centering
     \includegraphics[width=\linewidth]{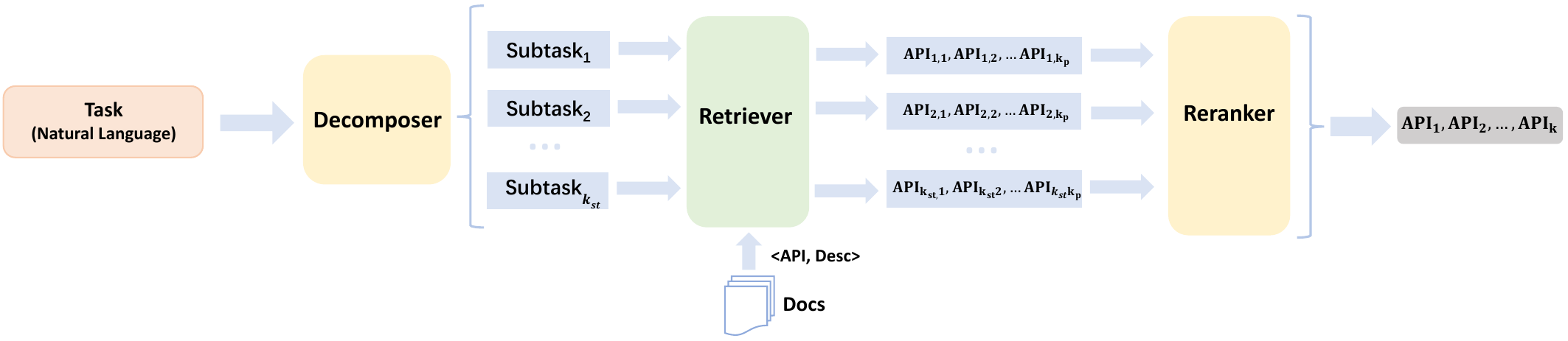}
     \caption{Overview of our compositional API recommendation.}
     \label{fig:framework}
 \end{figure*}

\section{Motivating Example} \label{sec:motivation}
Figure \ref{fig:noPrivateLibrary} shows the importance of API recommendation for LLMs and the limitation of raw-requirement-based API recommendation.
The development task  is \textit{"Load and only parse CSV files in the folder using Torchdata"}, which needs to use four APIs in \textit{Torchdata}: \textit{"FileLister"}, \textit{"Filter"}, \textit{"FileOpener"} and \textit{"parse\_csv"}. As shown in Figure \ref{fig:motivation1}, when the prompt of LLM does not include the corresponding APIs and their descriptions, LLM can only generate code that looks reasonable but cannot be executed correctly. 


The example in Figure \ref{fig:motivation1} illustrates the importance of including API information in the prompt, but the key challenge lies in accurately identifying the required APIs.
The task description can be coarse-grained, making it difficult to directly retrieve APIs from documents. 
As shown in figure \ref{fig:motivation2}, directly using the raw-requirement to retrieve from API documentation only yields \textit{"parse\_csv"}, which makes LLMs unable to generate the correct code.

To address this, we propose CAPIR, which decomposes the task into subtask sequence and uses these subtasks to retrieve APIs from documents. For the task "\textit{Load and only parse CSV files in the folder using Torchdata}", we decompose it into four subtasks: \{\textit{"List all files in the folder", "Filter out non-CSV files", "Open the CSV files", "Parse the CSV files"}\}. Then, we can retrieve corresponding APIs from the API documentation for each subtask: \{\textit{"FileLister", "Filter", "FileOpener", "parse\_csv"}\}. As shown in Figure \ref{fig:motivation3}, after adding the APIs recommended by CAPIR into the prompt, LLMs can generate the correct code.
Previous work has demonstrated the impressive reasoning capabilities of LLMs~\cite{kojima2022large, wei2022chain, wang2022self, li2023making}, so we utilize LLMs to decompose tasks into subtasks. Given that LLMs lack API-level granularity information, a critical challenge arises in determining the appropriate granularity for this decomposition process. The input length limitations of LLMs prevent us from providing an entire library documentation as input. We use in-context learning~\cite{dong2022survey} to address this issue. Low-resource libraries typically include API documentation with usage examples. We capitalize on these examples in the API documentation to construct task decomposition examples, effectively guiding the language model to decompose the task into API-level subtasks.

\section{Approach}

In this section, we will provide a detailed explanation of CAPIR approach. Figure~\ref{fig:framework} illustrates the pipeline of CAPIR, which consists of three main components: (1) \textbf{Task Decomposition}: CAPIR utilizes the LLM-based \textit{Decomposer} to break down a coarse-grained task into subtasks through few-shot prompting
(Section \ref{subsec:decomposer}). (2) \textbf{Subtask API Retrieval}: 
For each subtask, CAPIR employs the embedding-based \textit{Retriver} to search relevant APIs from the API documentation. (Section \ref{subsec:retriever}). (3) \textbf{API Reranking}: 
With the APIs retrieved from all subtasks, CAPIR leverages an LLM-based \textit{Reranker} to provide the final recommendation results.(Section \ref{subsec:reranker}).

    

\subsection{Task Decomposition}\label{subsec:decomposer}
CAPIR employs the LLM as a \textit{Decomposer} to effectively decompose coarse-grained development tasks into several subtasks. LLMs have already demonstrated powerful reasoning capabilities in various tasks~\cite{wei2022chain, chen2021evaluating}. The key challenge is \textit{how to demonstrate the granularity of task decomposition to the language model}. As previous work has shown the in-context learning~\cite{dong2022survey, bareiss2022code, brown2020language, an2023incontext} ability of LLMs, we use few-shot examples to demonstrate the decomposition granularity. 

\subsubsection{\textbf{Examples Construction}}
CAPIR constructs <task, subtasks> pairs as the decomposition examples. CAPIR collects code snippets that invoke multiple APIs, generates task descriptions for the code, and subtask descriptions for each API.
Due to the lack of resources for low-resource libraries on the internet, CAPIR leverages the usage examples in API documentation to construct example bank. The process contains three steps: First, CAPIR crawls API documentation of library $L_o$, denoted as $D_o$. Second, it extracts code examples that call more than two APIs from $D_o$. 
Third, it utilizes a Large Language Model $\mathbf{LLM_s}$ as \textit{Summarizer} to generate code summarization and the API function descriptions. The summarization process is denoted by the equation:
\begin{equation}\label{equ:example_summary}
<s_i,\mathcal{S}_i> =  \mathbf{LLM_s}(c_i,\mathcal{A}_i),
\end{equation}
where $s_i$ corresponds to the summarization of code snippet $c_i$, $\mathcal{A}_i$ refers to the API sequence that used by $c_i$, and $\mathcal{S}_i = \{st_1,...,st_{m_i}\}$ refers to the functional descriptions of the APIs called by $c_i$. We consider summarization $s_i$ as the task corresponding to example code $c_i$ and treat the functionality description $\mathcal{S}_i$ as the API-level subtasks. $\mathbf{LLM_s}$ refers to a large language model with specific prompt to generate <task, subtasks> pairs.
\begin{exmp}{Prompt For \textit{Summarizer}}{example_summarizer}
    \small
I will give you a code snippet that uses multiple APIs from library $L_o$. You need to write a summarization for the code as task description, and the subtasks that each one refers to an API in $L_o$ in the following format:\\
\{"task": task description, "subtasks": [subtask1, subtask2...] \}\\
\textit{Code snippet}: $c_i$\\
\textit{APIs}: $\mathcal{A}_i$\\
\textit{Task and subtasks}:
\end{exmp}

\subsubsection{\textbf{Examples Selection}}\label{subsec:example_selection}

Due to the input length limitation of LLM, CAPIR selects $k_e$ most relevant examples to add into the prompt. For a given task $t_i$, CAPIR adopts an embedding-based \textit{Selector} to select examples based on their functional similarity to $t_i$. In contrast to previous approaches~\cite{wei2022clear}, where the embedding model is fine-tuned using training data. As low-resource libraries do not have enough resources for training, CAPIR employs an off-the-shelf embedding model. This choice results a more efficient and effective \textit{Selector}. The \textit{Selector} has three steps: First, it calculates the embedding feature of the task description $t_i$ and the code summarization $s_j$ of example $c_j$. Second, it calculates the cosine similarity based on the embeddings. Third, it selects the $k_e$ most similar examples as the few-shot examples for \textit{Decomposer}. The cosine similarity is computed using the following formula:

\begin{equation}\label{equ:cosine_similarity}
sim(t_i,s_j) = \frac{\mathbf{Emb}(t_i) \cdot \mathbf{Emb}(s_j)}{\|\mathbf{Emb}(t_i)\| \|\mathbf{Emb}(s_j)\|}, 
\end{equation} where $\mathbf{Emb}$ represents the embedding model, $sim(t_i,s_j)$ quantifies the cosine similarity between the embeddings of task $t_i$ and summarization $s_j$.

\subsubsection{\textbf{Decomposer}} CAPIR employs  an LLM-based \textit{Decomposer} to decompose the task into several subtasks.
For a given task $t_i$, CAPIR prepares the input for the \textit{Decomposer} by including $k_e$ relevant examples, alongside the task $t_i$ itself. 
The output of this process is a subtask sequence.
Prompt \ref{exmp:example_decomposer} shows the \textit{Decomposer} prompt. 
The output of \textit{Decomposer} is the decomposed subtask sequence.
Formally, the decomposition process is expressed as:

\begin{equation}
\mathcal{S}_i = \mathbf{LLM_d}(\mathcal{E}_i,t_i),
\end{equation}
where $\mathbf{LLM_d}$ represents the \textit{Decomposer}, $\mathcal{S}_i = \{st_1,...,st_{m_i}\}$ denotes the subtask sequence, $\mathcal{E}_i = \{<s_1,\mathcal{S}_1>,...,<s_{k_e},\mathcal{S}_{k_e}>\}$ corresponds  to the few-shot examples for task $t_i$. 

\begin{exmp}{Prompt For \textit{Decomposer}}{example_decomposer}
    \small
    I will give you a task that needs to use several APIs to implement it. You need to break down the task into several subtasks. \\
    \textit{Examples}: $\mathcal{E}_i$\\
    \textit{Task}: $t_i$\\
    \textit{Subtasks}:
\end{exmp}

\subsection{Subtask API Retrieval}\label{subsec:retriever}

After decomposing the task $t_i$ into subtasks $\mathcal{S}_i$, we need to retrieve the corresponding APIs for each subtask. For a given subtask $st_j$, CAPIR uses \textit{Retriever} module to retrieve API from the API documentation $D_o$. Due to limited software resources related to a specific low-resource library, fine-tuning a library-specific embedding model is not feasible. CAPIR utilizes the same off-the-shelf embedding model as \textit{Selector}. The \textit{Retriever} also has three steps: First, it calculates the embedding feature of the subtask description $st_j$ and the description $desc_q$ of $API_q$. Second, it calculates the cosine similarity based on the embeddings. Third, it returns the $k_p$ most similar APIs, represented as $\mathcal{A}_j = {<API_1,desc_1>,...,<API_{k_p},desc_{k_p}>}$, as the API candidates for subtask $\mathbf{st_j}$. The cosine similarity between a subtask description and an API description is computed using the following formula:

\begin{equation}
sim(st_j, desc_q) = \frac{\mathbf{Emb}(st_j) \cdot \mathbf{Emb}(desc_q)}{\|\mathbf{Emb}(st_j)\| \|\mathbf{Emb}(desc_q)\|},
\end{equation}
where $\mathbf{Emb}$ represents the embedding model, $sim(st_j, desc_q)$ quantifies the cosine similarity between the embeddings of subtask $st_j$ and API description $desc_q$.

\subsection{API Reranking}\label{subsec:reranker}
CAPIR's LLM-based API reranking module first reranks the candidate APIs within subtasks, then reranks across subtasks to form the final recommended results.
There may be differences in the way subtasks and APIs are expressed, which can impact the API retrieval results. To address this, CAPIR utilizes a large language model as \textit{$Reranker_{ws}$} to rerank the retrieval results within subtasks. 
Meanwhile, the granularity of subtasks and APIs may not always be perfectly aligned, resulting in some subtasks corresponding to multiple APIs, while others may not require API implementation. To get the final compositional APIs for the task, CAPIR utilizes 
\textit{$Reranker_{as}$} to rerank across subtasks to get top-$k$ APIs.

\subsubsection{\textbf{Reranking within subtasks}} CAPIR reranks the candidate APIs within subtasks. 
By leveraging the comprehensive software knowledge and language comprehension capabilities of LLM, the module performs re-ranking of the APIs based on their relevance to the specific subtask.  
Prompt \ref{exmp:example_rerankerws} shows the \textit{$Reranker_{ws}$} prompt. 
Formally, the \textit{$Reranker_{ws}$} process can be expressed as follows:
\begin{equation}
\mathcal{P}_j = \mathbf{LLM_{rw}}(st_j,\mathcal{A}_j),
\end{equation}
where $\mathbf{LLM_{rw}}$ denotes the \textit{$Reranker_{ws}$} , and $\mathcal{P}_j = \{<API'_1,desc'_1>,...,<API'_{k_p},desc'_{k_p}>\}$ refers to the re-ordered API sequence.
The refined API sequence $\mathcal{P}_j$ provides more suitable and relevant APIs for developers.

\begin{exmp}{Prompt For \textit{$Reranker_{ws}$}}{example_rerankerws}
    \small
    I would like to use library $L_o$ to implement the subtask $st_j$. I searched for $k_p$ APIs from the documentation of $L_o$. Please rerank the APIs based on their probability of being able to solve the subtask.\\
    \textit{Subtask}: $st_j$\\
    \textit{Retrieved APIs}: $\mathcal{A}_j$\\
    \textit{Re-ordered APIs}:
\end{exmp}

\subsubsection{\textbf{Reranking across subtasks}} 
Since the decomposed subtasks cannot guarantee a one-to-one correspondence with APIs, directly combining the retrieved APIs for each subtask might result in the omission of necessary APIs or the inclusion of redundant ones. CAPIR leverages cross-subtask API reranker \textit{$Reranker_{as}$} to select APIs from each subtask's retrieval results to form the final API recommendation. 
Prompt \ref{exmp:example_rerankeras} shows the \textit{$Reranker_{as}$} prompt. 
Formally, the \textit{$Reranker_{as}$} process can be expressed as follows:

\begin{equation}
\mathcal{I}_i = \mathbf{LLM_{ra}}(s_i,\mathcal{S}_i, \mathcal{R}_i),
\end{equation}
where $\mathbf{LLM_{ra}}$ denotes the \textit{$Reranker_{as}$} model, $\mathcal{R}_i = \{\mathcal{P}_1,...,\mathcal{P}_{m_i}\}$ refers to the re-ordered API sequence corresponding to subtask sequence $\mathcal{S}_i = \{st_1,...,st_{m_i}\}$, and the result $\mathcal{I}_i = \{<API''_1,desc''_1>,...,<API''_{k},desc''_{k}>\}$ refers to the final recommended API sequence.

\begin{exmp}{Prompt For \textit{$Reranker_{as}$}}{example_rerankeras}
    \small
    I would like to use library $L_o$ to implement task $t_i$. I have decompose the task into subtasks and searched $k_p$ APIs for each subtask from the documentation. \\
    I hope to combine the APIs corresponding to each subtask to get the API sequence to solve the task. Please give me the top-$k$ APIs that sorted based on their probability of being able to solve the task.\\
    \textit{Task}: $t_i$\\
    - \textit{Subtask 1}: $st_1$\\
    - \textit{Retrieved APIs for subtask 1}: $\mathcal{P}_1$\\
    ...\\
    - \textit{Subtask $m_i$}: $st_{m_i}$\\
    - \textit{Retrieved APIs for subtask $m_i$}: $\mathcal{P}_{m_i}$\\
    \textit{Final top-$k$ APIs}:
\end{exmp}

\begin{table}
\caption{Dataset Statistics}
\label{tab:datasets}
\resizebox{\linewidth}{!}{\begin{tabular}{cccccc}
\hline
Benchmark & Dataset & \# samples & Avg. APIs & Max. APIs & \# Candidate APIs\\
\hline
\multirow{5}{*}{RAPID} &
Torchdata-AR & 278 & 4.35 & 8 & 96 \\
&Math-AR & 200 & 2.20 & 8 & 88 \\
&RealMatrix-AR & 48 & 2.54 & 5 & 26 \\
&XMLStreamWriter-AR & 187 & 3.20 & 8 & 26 \\
&Multi-Conala & 232 & 2.48 & 7 & 30775 \\
\hline
\multirow{2}{*}{LOCG} &
Torchdata-Code & 50 & 4.22 & 8 & 96 \\
&Multi-ODEX & 67 & 2.17 & 3 & 30775 \\
\hline
\end{tabular}}

\end{table}




\section{EXPERIMENTAL SETUP}
In this section, we will introduce our experimental setup, including the research questions, models, benchmarks, baselines, and evaluation metrics.

\subsection{Research Questions}
To evaluate the performance of CAPIR on API sequence recommendation and library-oriented code generation, we conducted experiments to address the following research questions:

\textbf{RQ1}: How effective is CAPIR for Documentation-based API sequence recommendation?

\textbf{RQ2}: How effective are the APIs recommended by CAPIR in library-oriented code generation?

\textbf{RQ3}: How effective are the individual components of CAPIR in improving performance?

\textbf{RQ4}: How effective is CAPIR with different large language models?

In RQ1, we compare CAPIR with baselines on API sequence recommendation for both single-library and multi-libraries. In RQ2, we compare CAPIR with baselines on enhancing library-oriented code generation for both single-library and multi-libraries. In RQ3, we compare CAPIR with three ablation settings: (1) Without Examples. We remove the examples in the Decomposer prompt, to assess whether they have positive impact on CAPIR. (2) Without Decomposer. We remove the Decomposer module and directly use the original task to retrieve APIs from Docs, while keeping the Reranker module, to assess whether the Decomposer module improves CAPIR's performance. (3) Without Reranker. We remove the Reranker module and directly return the top-k retrieve results, to assess whether the Reranker can improve the retrival performance. In RQ4, we evaluate the API recommendation performance of CAPIR with another LLM.

\subsection{Models}

For the Summarizer, Decomposer, and Reranker, we employ the \textit{gpt-3.5-turbo}~\cite{chatgpt} model, which has demonstrated exceptional performance in various natural language processing tasks. The prompt for these modules is tailored to their specific tasks. To ensure the certainty of the outputs, we set the parameters of \textit{gpt-3.5-turbo} as follows: \textit{temperature}=0.0, \textit{top\_p}=1.0. For Decomposer, we set few-shot examples number $k_e$ = 4, to balance the example diversity and the sequence length. For Reranker, we set candidate API number $k_p$ = 20. 

On the other hand, for the Selector and Retriever modules, we utilize the \textit{ada-embedding-002}~\cite{ADA-embedding-002} model as the embedding model. This model has the ability to generate high-quality embeddings for sentences, allowing us to perform effective similarity computations and retrieval. 

\subsection{Benchmark Construction}
To evaluate API sequence recommendation and library-oriented code generation for coarse-grained developmental requirements, we construct two challenging benchmarks: (1) RAPID, a documentation-based API sequence recommendation benchmark. (2) LOCG, a library-oriented code generation benchmark.

\subsubsection{\textbf{RAPID}} The RAPID benchmark contains five datasets. Four single-library API recommendation datasets that correspond to four libraries: Torchdata-AR (AR refers to API Recommendation), Math-AR, Real-Matrix-AR and XMLStreamWriter-AR. One multi-libraries API recommendation task: Multi-Conala, contains seven libraries.
The statistics for RAPID are shown in Table \ref{tab:datasets}.

\textbf{Torchdata-AR}. To evaluate CAPIR for low-resource libraries, we use Torchdata to construct Torchdata-AR datasets. 
Public information shows that gpt-3.5-turbo was trained on data collected from the internet before September 2021. Since Torchdata was released on Mar 11, 2022, it was not included in the training data of gpt-3.5-turbo. Moreover, it has been released for more than one year, ensures there is sufficient client code in the open-source community to construct a test set. Construction details of Torchdata-AR will be provided in Section~\ref{subsec:RAPID}. 

\textbf{Math-AR, RealMatrix-AR and XMLStreamWriter-AR}. 
These three datasets are constructed based on DGAS~\cite{wei2023documentation}, an API recommendation dataset. The original DGAS dataset includes training data and documentations corresponding to these libraries. In the datasets we constructed, we excluded the training data and only retained test samples that refer to multiple APIs.

\textbf{Multi-Conala}. To validate the effectiveness of our approach in API sequence recommendation for multi-libraries, we conducted experiments using the well-established Conala dataset~\cite{yin2018learning}. Conala contains a diverse and representative set of popular libraries, making it a valuable resource for conducting experiments in the field of API recommendation. Following the methodology outlined in a previous study ~\cite{zhou2022docprompting}, we obtained API documentation from DevDocs\footnote{https://devdocs.io}, encompassing a total of 30,755 APIs extracted from different libraries.
To evaluate the API sequence recommendation for coarse-grained requirements, we filtered out samples that only refer to one API. This filtering step resulted in a new test dataset of 232 samples and a training set of 625 samples. These samples ensure a diverse range of API recommendation scenarios, allowing us to comprehensively evaluate the performance of our approach on API recommendation for multi-libraries.

\subsubsection{\textbf{LOCG}} The LOCG benchmark contains two datasets: (1) Torchdata-Code, a single-library-oriented code generation dataset. (2) Multi-ODEX, a multi-library-oriented code generation dataset.
The statistics of LOCG are shown in Table \ref{tab:datasets}.

\textbf{Torchdata-Code}. To evaluate the impact of recommended APIs on single-library-oriented code generation performance, we selected 50 samples from  the Torchdata-AR to construct an excution-based code genration dataset. We extracted the original code snippets, manual transformed  them into executable functions and wrote test cases for them.

\textbf{Multi-ODEX}. To evaluate the impact of recommended APIs on multi-library-oriented code generation performance, we constructed code generation dataset for multi-libraries based on ODEX~\cite{wang2022execution} dataset. ODEX is a benchmark for evaluating natural language (NL) to Python code generation, as it extracts coding queries from StackOverflow, representing practical NL-to-code scenarios.
To ensure a focused evaluation, we filtered out examples in the ODEX dataset that involved calling only one API. This filtering step resulted in a carefully curated test set comprising 67 samples. Since ODEX is constructed based on conala, and the APIs they call belong to the same libraries, we used the same API documentation as in Multi-Conala.


\subsubsection{\textbf{Torchdata-AR Construction Details}} \label{subsec:RAPID}
We crawled repositories on GitHub that import the Torchdata library to construct the Torchdata API recommendation dataset. By using the keywords "from\ Torchdata" and "import\ Torchdata," we utilized the Github Search code's REST API and collected 1342 code files. We first removed duplicates and split  the code files at the function granularity. Then, we filtered out code snippets that called fewer than 3 APIs.
Since most of these code snippets lack descriptive comments, we need to summarize the functionality of the code snippets to use them as task descriptions during testing. To reduce manual annotation costs, we use GPT-4~\cite{openai2023gpt4} to generate comments for these code snippets. GPT-4 has achieved comparable to or even surpassing human performance in many tasks, making it competent for our annotation. For each code snippet, we extracted the Torchdata APIs that were  used in the code and added  their API descriptions into the prompt, enabling GPT-4 to generate higher quality code comments. 

To enhance the credibility and quality of the data generated by GPT-4, we engaged three graduate students with more than five years of programming experience to scrutinize the comments generated by GPT-4. We first let them thoroughly read the Torchdata documentation to familiarize themselves with the functionalities and usage of the APIs. Then the code comments generated by GPT-4 were carefully examined, and were categorized into three types: "appropriate," "require improvement," and "inappropriate."

1) For samples that all three annotators deemed "appropriate", they were included in the final test set without any modifications.

2) If the comment is marked as "require improvement", the annotators engaged in discussions to reach a consensus on how to improve it. If a consensus was reached, the sample was added to the test set after the new comment replaced  the original one. On the other hand, if no agreement could be reached, the sample was excluded from the test set.

3) Comments that were found to be "inappropriate" by any of the annotators were immediately discarded from the dataset to ensure its high quality and reliability.

Following this rigorous curation process, we finalized the Torchdata API recommendation dataset, comprising a total of 278 examples. This curated test set 
ensures  a robust evaluation of our approach on realistic development scenarios.

\subsection{Baselines}\label{sec:baselines}
We compare CAPIR with two baselines: (1) ADA-retrieve, which directly employs CAPIR's base embedding model for retrieval.(2) CLEAR~\cite{wei2022clear}, the state-of-the-art retrieval-based API recommendation method.
\begin{itemize}[leftmargin=*]
    \item \textbf{ADA-retrieve} ~\cite{ADA-embedding-002}. To demonstrate the effectiveness of task decomposition for retrieval, we chose ADA-retrieve as a baseline. ADA-retrieve directly used OpenAI's off-the-shelf embedding model, ada-embedding-002, to retrieve the top-$k$ APIs from the API documentation based on the task description's similarity to the functional descriptions.
    \item  \textbf{CLEAR} ~\cite{wei2022clear}. CLEAR is the state-of-the-art method for API recommendation tasks. It embeds queries and Stack Overflow posts using BERT to preserve sequential information and then employs contrastive learning to learn precise semantic representations of programming terminologies. It also incorporates a BERT-based re-ranking model to optimize recommendation results. 
\end{itemize}

\begin{table*}
    \centering
    \caption{Performance comparison on API Recommendation. The first four are single-library API sequence recommendation results, and the last one is multi-library API sequence recommendation results.}
    \label{tab:APIRec}
    \begin{tabular}{cccccccccc}
    \hline
    \multirow{2}{*}{Dataset} & \multirow{2}{*}{Method} & \multicolumn{2}{c}{\bfseries{$k=3$}} & \multicolumn{2}{c}{\bfseries{$k=5$}} & \multicolumn{2}{c}{\bfseries{$k=10$}} & \multicolumn{2}{c}{\bfseries{$k=15$}} \\
    \cmidrule(lr){3-4} \cmidrule(lr){5-6} \cmidrule(lr){7-8} \cmidrule(lr){9-10}
    && Recall & Precision & Recall & Precision & Recall & Precision & Recall & Precision \\
    \hline
    \multirow{3}{*}{Torchdata-AR} 
    &ADA-retrieve & 17.6\% & 25.1\% & 23.6\% & 20.1\% & 34.3\% & 14.8\% & 43.7\% & 12.6\% \\
    &CLEAR & 13.2\% & 18.4\% & 18.7\% & 15.5\% & 29.2\% & 12.2\% & 39.6\% & 11.1\% \\
    &  \colorgray CAPIR (ours) &  \colorgray\textbf{30.6\%} & \colorgray\textbf{43.7\%} & \colorgray\textbf{43.2\%} & \colorgray\textbf{37.1\%} & \colorgray\textbf{54.4\%} &  \colorgray\textbf{23.4\%} &  \colorgray\textbf{60.4\%} &  \colorgray\textbf{17.3\%}\\
    \hline
    \multirow{3}{*}{Math-AR} 
    &ADA-retrieve & 16.0\% & 11.2\% & 26.8\% & 11.2\% & 46.7\% & 9.8\% & 58.2\% & 8.2\% \\
    &CLEAR & 11.3\% & 9.0\% & 15.3\% & 9.1\% & 18.6\% & 8.9\% & 39.6\% & 6.6\% \\
    &  \colorgray CAPIR (ours) & \colorgray\textbf{31.6\%} & \colorgray\textbf{23.5\%} & \colorgray\textbf{42.8\%} & \colorgray\textbf{19.4\%} & \colorgray\textbf{55.7\%} & \colorgray\textbf{15.3\%} & \colorgray\textbf{64.1\%} & \colorgray\textbf{9.5\%}\\
    \hline
    \multirow{3}{*}{RealMatrix-AR} 
    &ADA-retrieve & 15.0\% & 12.5\% & 22.9\% & 12.1\% & 37.1\% & 9.6\% & 57.1\% & 9.7\% \\
    &CLEAR & \textbf{27.3\%} & 20.7\% & 30.0\% & 11.0\% & 33.9\% & 9.6\% & 57.7\% & 6.8\% \\
    &  \colorgray CAPIR (ours) & \colorgray{26.4\%} & \colorgray\textbf{22.2\%} & \colorgray\textbf{35.0\%} & \colorgray\textbf{18.3\%} & \colorgray\textbf{49.3\%} & \colorgray\textbf{15.9\%} & \colorgray\textbf{63.6\%} & \colorgray\textbf{11.1\%}\\
    \hline
    \multirow{3}{*}{XMLStreamWriter-AR} 
    &ADA-retrieve & 20.7\% & 23.0\% & 28.8\% & 19.3\% & 58.1\% & 18.9\% & 79.2\% & 17.0\% \\
    &CLEAR & 36.0\% & 41.2\% & 46.9\% & 31.3\% & 54.7\% & 19.8\% & 78.2\% & 16.8\% \\
    &  \colorgray CAPIR (ours) & \colorgray\textbf{43.3\%} & \colorgray\textbf{46.0\%} & \colorgray\textbf{55.3\%} & \colorgray\textbf{35.3\%} & \colorgray\textbf{63.7\%} & \colorgray\textbf{24.5\%} & \colorgray\textbf{83.4\%} & \colorgray\textbf{17.6\%}\\
    \hline
    \multirow{3}{*}{Multi-Conala} 
    &ADA-retrieve & 7.9\% & 6.5\% & 10.6\% & 5.3\% & 15.5\% & 3.8\% & 21.8\% & 3.6\% \\
    &CLEAR & 23.2\% & 19.2\% & 28.2\% & 14.4\% & 32.6\% & 8.4\% & 34.3\% & 5.9\% \\
    &  \colorgray CAPIR (ours) & \colorgray\textbf{30.0\%} & \colorgray\textbf{24.9\%} & \colorgray\textbf{35.3\%} & \colorgray\textbf{17.5\%} & \colorgray\textbf{39.7\%} & \colorgray\textbf{9.9\%} & \colorgray\textbf{42.3\%}& \colorgray\textbf{7.0\%}\\
    \hline
    \end{tabular}
    \end{table*}

\subsection{Metrics}
In this section, we will separately introduce the evaluation metrics for API recommendation and code generation.
To evaluate the performance of our API recommendation approach, we adopt standard evaluation metrics commonly used in existing studies~\cite{wei2022clear}: $Precision@k$ and $Recall@k$. where $k=\{3, 5, 10, 15\}$. To evaluate the impact of recommended APIs on library-oriented code generation, we use the execution based metric $pass@k$ to assess the accuracy of the generated code, where $k = \{1,10, 100\}$.

\textbf{Precision@k} measures the proportion of correctly recommended APIs among the top-k recommended APIs. It assesses the accuracy of our approach in providing relevant and correct API recommendations to developers.

\begin{equation}
    Precision@k= \mathcal{C}(\{retrieved\ APIs\}@k, \{relevant\ APIs\}),
\end{equation}
where 
\begin{equation}
    \mathcal{C}(A,B) = \frac{\sum_{a \in A} \delta(a, B)}{|A|},
\end{equation}
\begin{equation}
    \delta(a, B) = \begin{cases} 1 & \text{if } a \in B \\ 0 & \text{otherwise} \end{cases}.
\end{equation}
\textbf{Recall@k}, on the other hand, measures the proportion of correctly recommended APIs out of all ground-truth APIs within the top-k recommendation results. It evaluates the ability of our approach to retrieve a sufficient number of relevant APIs, ensuring that no important API is overlooked.
\begin{equation}
    Recall@k= \mathcal{C}(\{relevant\ APIs\}, \{retrieved\ APIs\}@k). 
\end{equation}
By calculating Precision@k and Recall@k for various values of k, we can comprehensively assess the effectiveness and coverage of our API recommendation approach. 

\textbf{Pass@k} is an evaluation metric that has been widely used in previous work~\cite{zan2022cert, zan2022language, chen2022codet}. Each task in LOCG has several corresponding test cases. For a given task, we randomly sample k codes and calculate the proportion that can pass all test cases.




\section{Results and Analyses}
In this section, we will present the results for the four RQs, analyze qualitative examples, and evaluate the performance of CAPIR in practical development scenarios.

\subsection{RQ1: Effectiveness of CAPIR on API Sequence Recommendation}

\textbf{Setup}. To compare the performance of CAPIR with two baselines in API sequence recommendation, we conducted experiments on the RAPID benchmark. Due to the limited posts about low-resource library in StackOverflow and other question-answering communities, we were unable to follow the exact method used in the CLEAR paper to select APIs from similar StackOverflow posts. Instead, we employed CLEAR to retrieve the closest $k$ APIs from the API documentation as the results. 

\textbf{Results}. The results from Table \ref{tab:APIRec} demonstrate that CAPIR outperforms the two baselines in compositional  API recommendation. 
(1) \uline{CAPIR outperforms the two baselines on all four single library API sequence recommendation datasets.} For Torchdata-AR dataset, comparing with ADA-retrieve, CAPIR improves the \textit{Recall}@3, 5, 10, 15 by 73.9\%, 83.1\%, 58.6\%, 38.2\% and \textit{Precision}@3, 5, 10, 15 by 74.1\%, 84.6\%, 58.1\%, 37.3\%; comparing with CLEAR, CAPIR improves the \textit{Recall}@3, 5, 10, 15 by 131.8\%, 131.0\%, 86.3\%, 52.5\% and \textit{Precision}@3, 5, 10, 15 by 137.5\%, 139.4\%, 91.8\%, 55.9\%. CAPIR also outperforms the two baselines on Math-AR, RealMatrix-AR and XMLStreamWriter-AR datasets. 
(2) \uline{CAPIR also outperforms the two baselines on multi-libraries API sequence recommendation.} For Multi-Conala dataset, Comparing CAPIR with ADA-retrieve, CAPIR improves the \textit{Recall}@3, 5, 10, 15 by 279.7\%, 233.0\%, 156.1\%, 94.0\% and \textit{Precision}@3, 5, 10, 15 by 283.1\%, 230.2\%, 160.5\%, 94.4\%; Comparing CAPIR with CLEAR, CAPIR improves the \textit{Recall}@3, 5, 10, 15 by 29.3\%, 25.2\%, 21.8\%, 23.3\% and \textit{Precision}@3, 5, 10, 15 by 29.7\%, 21.5\%, 17.9\%, 18.6\%. CAPIR's performance on Multi-Conala demonstrates that it can recommend more appropriates APIs  from a pool of multiple libraries.
Under most experimental settings, CAPIR achieves higher performance, which indicates better API sequence recommendation performance. Morevover, the comparison between CAPIR and CLEAR further illustrates  that previous methods perform less effectively on unseen libraries.

\find{
{\bf Answer to RQ1:} 
CAPIR outperforms ADA-retrieve and CLEAR on RAPID benchmark. The significant improvements indicate the effectiveness of CAPIR on documentation-based API sequence recommendation.
}

\subsection{RQ2: Effectiveness of CAPIR on Library-Oriented Code Generation}

\textbf{Setup}. To evaluate the impact of CAPIR on library-oriented code generation, we conducted experiments on Torchdata-Code and Multi-ODEX datasets. We used gpt-3.5-turbo as the base model for code generation. 
The prompt for code generation is shown in \ref{fig:motivation3},  
including the development task and the APIs recommended by CAPIR. To calculate pass@k with multiple samples, as done in previous studies, we set the temperature = 0.8 and top\_p=0.95. The number of APIs included in the prompt is a crucial consideration. If there are too few APIs in the prompt, it may not cover the required APIs for the given task, while having too many APIs could introduce excessive noise to the prompt. According to the results in Table \ref{tab:APIRec}, k=5 strikes the best trade-off between recall and precision. Therefore, we add 5 APIs for each task in the prompt.

\begin{table}
\centering
\caption{Evaluation results on the library-oriented code generation.\label{tab:odex}}

\resizebox{\linewidth}{!}{\begin{tabular}{ccccc}
\hline
Dataset & Method & Pass@1 & Pass@10 & Pass@100\\
\hline
\multirow{3}{*}{Torchdata-Code} 
& GPT-3.5 & 1.7 & 9.1 & 16.0 \\
& GPT-3.5 + ADA-retrieve & 1.3 & 7.0 & 12.0 \\
& \colorgray GPT-3.5 + CAPIR (ours) & \colorgray \textbf{3.0} & \colorgray \textbf{12.4} & \colorgray \textbf{28.0}\\
\hline
\multirow{3}{*}{Multi-ODEX} 
& GPT-3.5 & 27.5 & 35.8 & 41.8 \\
& GPT-3.5 + ADA-retrieve & 27.6 & 38.8 & 44.8\\
& \colorgray GPT-3.5 + CAPIR (ours) & \colorgray \textbf{35.2} & \colorgray \textbf{44.8} & \colorgray \textbf{55.2} \\
\hline
\end{tabular}}
\end{table}

\textbf{Results}. The results of Table ~\ref{tab:odex} show that adding CAPIR's recommendation results into the prompt can improve code generation performance. 
(1) \uline{On single-library-oriented code generation.} Compared to  using gpt-3.5-turbo to directly generate code, adding CAPIR's recommendation results improves \textit{pass}@1, 10, 100 by 76.47\%, 36.26\%, 75.0\%. Comparing with adding ADA-retrieve results, adding CAPIR's recommendation results improves \textit{pass}@1, 10, 100 by 130.77\%, 77.14\%, 133.33\%. It's worth noting that on the Torchdata-Code dataset, adding ADA-retrieve results to the prompt decreased the code generation performance compared to directly generating code. This suggests that when the quality of API recommendations is insufficient, it will introduce noise, thereby having a negative impact on library-oriented code generation.
(1) \uline{On multi-library-oriented code generation.}
Compared to  using gpt-3.5-turbo to directly generate code, adding CAPIR's recommendation results improves \textit{pass}@1, 10, 100 by 28.0\%, 25.14\%, 32.06\%. Compared to  adding ADA-retrieve results, adding CAPIR's recommendation results improves \textit{pass}@1, 10, 100 by 27.54\%, 15.46\%, 23.21\%. 
CAPIR exhibits a more significant improvement on Torchdata-Code compared to Multi-ODEX, indicating that adding API information to the prompt is more effective for unseen-library-oriented code generation.
Overall, the results suggest that the quality of API recommendations plays a crucial role in enhancing the code generation effectiveness.

\find{
{\bf Answer to RQ2:}  Compared to ADA-retrieve, the API recommended by CAPIR can bring more improvements to library-oriented code generation.
}

\subsection{RQ3: Effectiveness of Key Components}

\textbf{Setup}. We conducted ablation experiments on Torchdata-AR dataset to assess the effectiveness of three crucial components of CAPIR: (1) CAPIR without the examples in the Decomposer prompt. (2) CAPIR without the Decomposer module, directly retrieves APIs from Docs, but remains the Reranker module. (3) CAPIR without Reranker. 

\begin{figure}
    \centering
    \subfloat[Recall vs. k]{
    \includegraphics[width=0.22\textwidth]{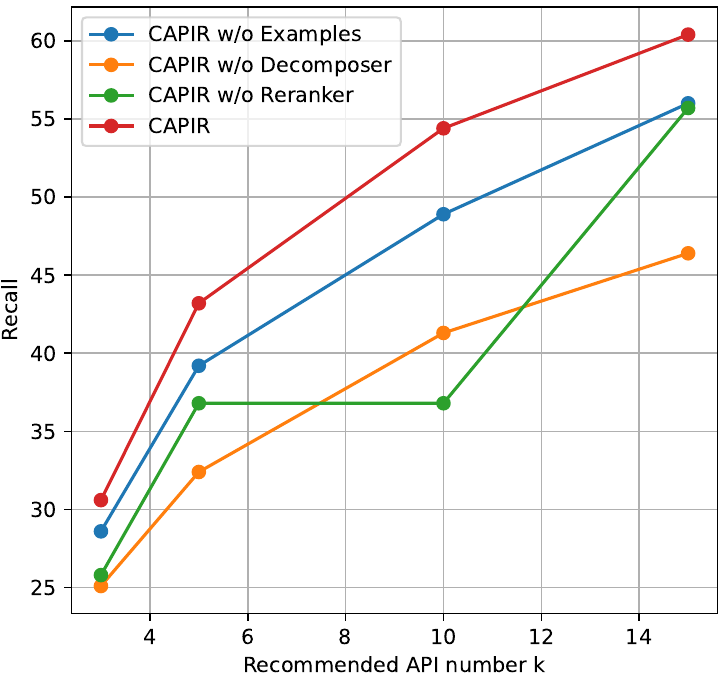}}\quad
    \subfloat[Precision vs. k]{
    \includegraphics[width=0.22\textwidth]{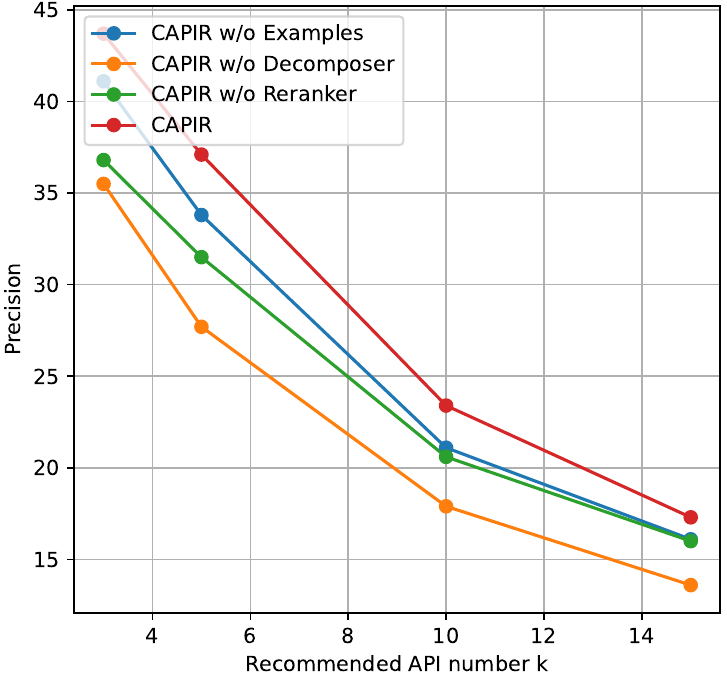}}\quad
    
    \caption{Recall@k and precision@k of CAPIR with and without key components on Torchdata-AR. Removing any part would lead to a decline in CAPIR's performance.}
    \label{fig:APIRec_ablation}
\end{figure}

\textbf{Results}. The results in Figure \ref{fig:APIRec_ablation} show that all three components play a crucial role for CAPIR performance. On Torchdata-AR dataset, adding the Examples improves CAPIR's \textit{Recall}@3, 5, 10, 15 by 7.0\%, 10.2\%, 11.2\%, 7.9\% and \textit{Precision}@3, 5, 10, 15 by 6.3\%, 9.8\%, 9.0\%, 7.5\%, adding the Decomposer improves CAPIR's \textit{Recall}@3, 5, 10, 15 by 21.9\%, 33.3\%, 31.7\%, 30.2\% and \textit{Precision}@3, 5, 10, 15 by 23.1\%, 33.9\%, 30.7\%, 27.2\%, adding the Reranker improves CAPIR's \textit{Recall}@3, 5, 10, 15 by 18.6\%, 17.4\%, 13.3\%, 8.4\% and \textit{Precision}@3, 5, 10, 15 by 18.8\%, 17.8\%, 13.6\%, 8.1\%. As k increases, the enhancement of Reranker on CAPIR decreases, suggesting that Reranker has a more significant impact when fewer APIs are recommended to users.
The results show that all the three components are crucial for achieving the best performance. The addition of the Decomposer yielded the most significant improvement, indicating that the Decomposer plays the most critical role in CAPIR's performance. 

\find{
{\bf Answer to RQ3:}  ALL three components of CAPIR are essential for its performance. The Decomposer plays the most critical role in CAPIR's performance.
}


\begin{figure*}
     \centering
     \includegraphics[width=\linewidth]{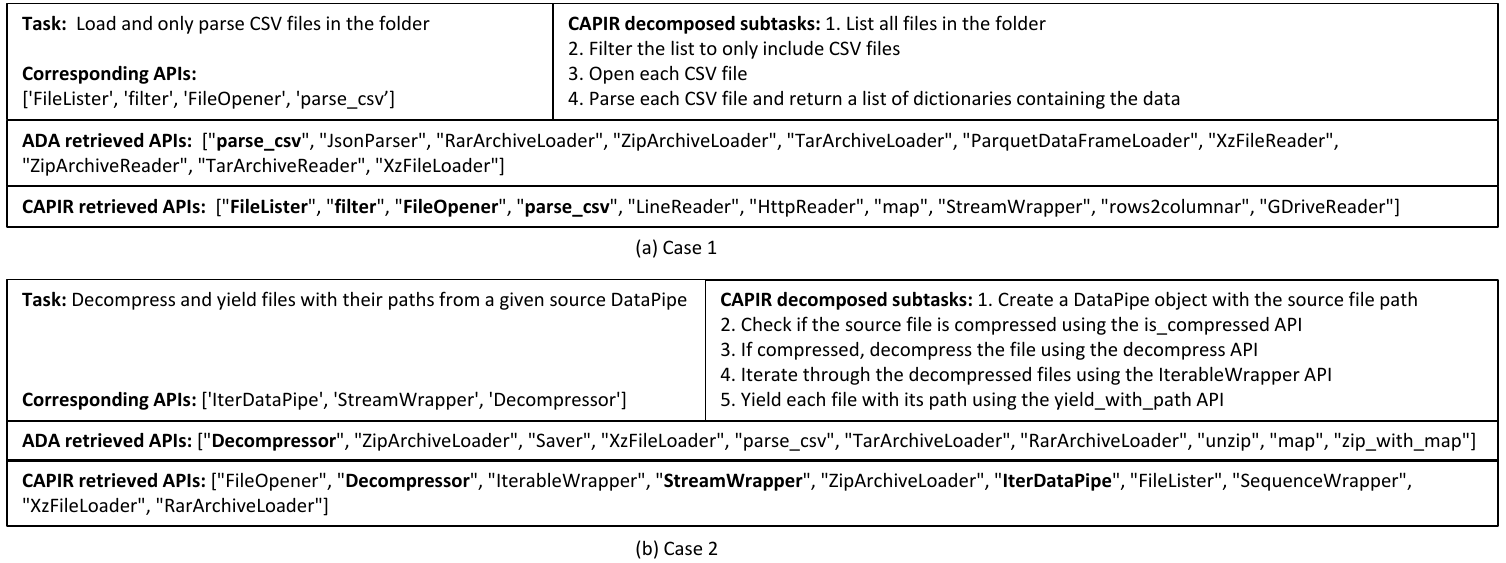}
     \caption{API recommendation results for two cases. (a) CAPIR decomposes the task into the appropriate granularity and effectively
recommends the correct APIs from the documentation. (b) CAPIR does not precisely decompose the task into subtasks of optimal granularity, but the decomposition still enhances the effectiveness of API recommendation.}
     \label{fig:case}
 \end{figure*}

\begin{table}
\centering

\caption{API recommendation results of CAPIR with different LLMs.\label{tab:otherLLM}}
\resizebox{\linewidth}{!}{\begin{tabular}{cccccccc}
\hline
Method & Recall@5 & Precision@5 & Recall@10 & Precision@10\\
\hline
ADA-retrieve & 23.6\% & 20.1\% & 34.3\% & 14.8\% \\
CLEAR & 18.7\% & 15.5\% & 29.2\% & 12.2\% \\
\colorgray CAPIR w/ llama-2-70b & \colorgray 36.9\% & \colorgray 31.4\% & \colorgray 48.1\% & \colorgray 20.5\% \\
\colorgray CAPIR w/ gpt-3.5-turbo & \colorgray \textbf{43.2\%} & \colorgray \textbf{37.1\%} & \colorgray \textbf{54.4\%} & \colorgray \textbf{23.4\%} \\
\hline
\end{tabular}}
\end{table}

\subsection{RQ4: Effectiveness of CAPIR with different LLMs}
\textbf{Setup.} To evaluate CAPIR's performance with different LLMs, we conducted an experiment to evaluate the API recommendation performance of CAPIR with another LLM. Specifically, we replaced the gpt-3.5-turbo used in CAPIR with llama-2-70b, and conducted experiments on the Torch-AR dataset.

\textbf{Results.} The results in Table \ref{tab:otherLLM} indicate that CAPIR still maintains a clear advantage with other LLMs. For Torchdata-AR dataset, comparing with ADA-retrieve, CAPIR with llama-2-70b improves the \textit{Recall}@5, 10 by 56.4\%, 40.2\% and \textit{Precision}@5, 10 by 56.2\%, 38.5\%; comparing with CLEAR, CAPIR with llama-2-70b improves the \textit{Recall}@5, 10 by 97.3\%, 64.7\% and \textit{Precision}@5, 10 by 102.6\%, 68.0\%. CAPIR with llama-2-70b and gpt-3.5-turbo both outperform the baselines, indicating its effectiveness across different LLMs. Compared to CAPIR with gpt-3.5-turbo, CAPIR with llama-2-70b performs less well. Since llama-2-70b is a base model without instruction tuning, its weaker ability to follow instructions has consequently impacted the performance of CAPIR.

\find{
{\bf Answer to RQ4:} CAPIR still maintains a clear advantage with other LLMs. The performance gap between CAPIR with llama-2-70b and CAPIR with gpt-3.5-turbo indicates that the ability of language models, such as instruction following ability, affects the performance of CAPIR.
}

\subsection{Qualitative Analysis}
In this section, we will conduct a more in-depth qualitative analysis of two examples. Figure \ref{fig:case} shows two API recommendation cases on Torchdata-AR dataset. From these examples, we obtain the following findings:

(1) Case 1 analysis. CAPIR showcases its remarkable capability to accurately decompose tasks into the appropriate granularity and effectively retrieve the correct APIs from the documentation. For the task "Load and only parse CSV files in the folder" mentioned in Section \ref{sec:motivation}, where the corresponding APIs should be ['FileLister', 'filter', 'FileOpener', 'parse\_csv']. CAPIR succeeds in decomposing tasks into four API-granular subtasks and places the corresponding APIs at the top of the retrieval results for each subtask. Finally, it combines the retrieved APIs into an API sequence and returns it to the user. In contrast, the ADA-retrieve results contain only "parse\_csv" within the top-10 ranked APIs. This example demonstrates that CAPIR is capable of decomposing tasks into API-level subtasks and accurately retrieving the corresponding APIs from the documentation.

\textbf{Further analysis for case 1}.
To further analyze why CAPIR outperforms the baseline, we conduct a more in-depth analysis on this example. First, we use the UMAP~\cite{mcinnes2018umap} algorithm to reduce the embeddings of APIs and tasks to 2-dimensions. Next, we create co-occurrence feature vectors for each API. For $API_i$, the j-th element represents the number of times $API_i$ and $API_j$ were combined to complete a task in test datasets. Then, we use the HDBSCAN~\cite{campello2013density} clustering algorithm to color all APIs based on their co-occurrence features.

\begin{figure}
     \centering
     \small
     \includegraphics[width=0.9\linewidth]{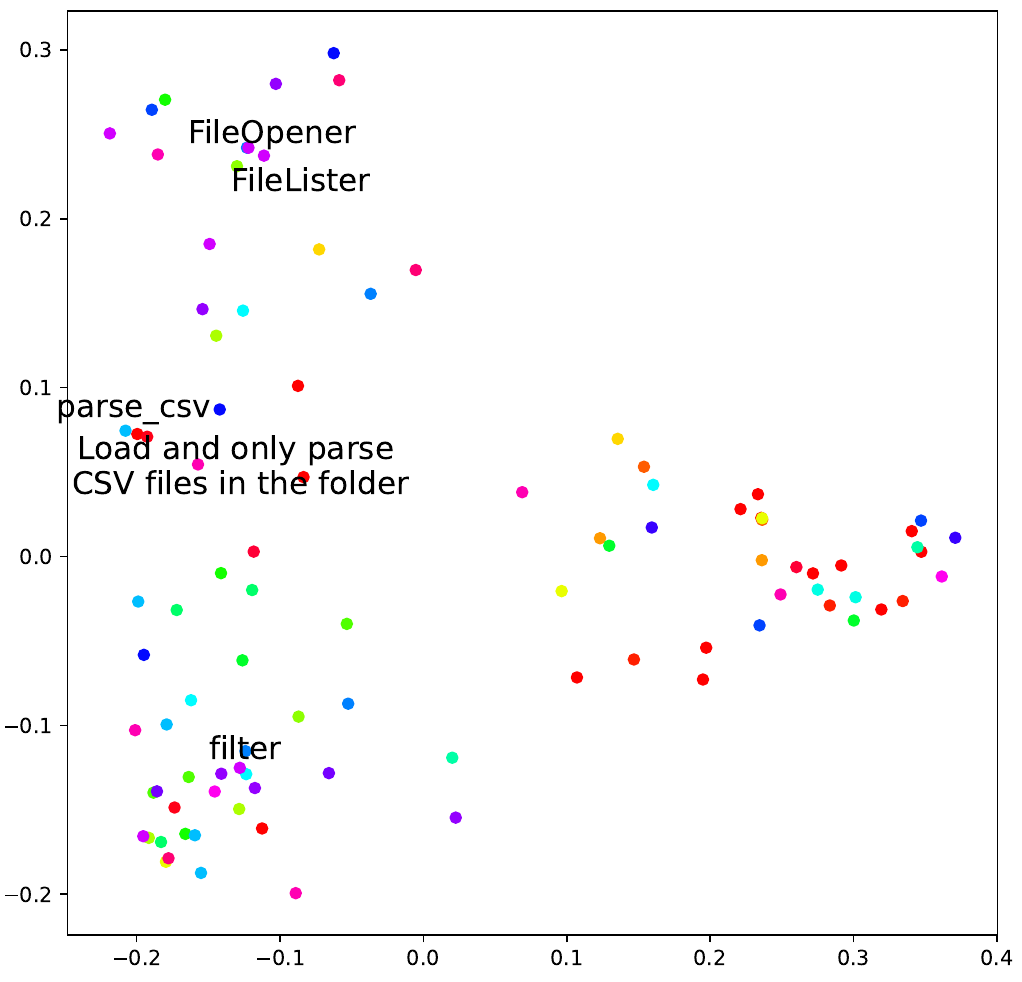}
     \caption{Visualization of all the API embeddings of Torchdata. The APIs corresponding to a coarse-grained task are scattered in different locations, indicates the necessity of decomposing tasks into subtasks for retrieval.}
     \label{fig:analysis}
      \vspace{-0.3cm}
 \end{figure}

From the visualization results in \ref{fig:analysis}, we can see that nodes of similar colors are scattered throughout the graph. The results means that APIs with high co-occurrence frequencies were not closely clustered in the embedding feature space, making it challenging to retrieve relevant multiple APIs based on a single task description. For instance, the four APIs corresponding to task "Load and only parse CSV files in the folder" are scattered throughout the graph, which makes ADA-retrieve can only retrieve "parse\_csv" based on the task description.
This analysis highlights the importance of CAPIR's approach in decomposing the task into several subtasks, which allows for more accurate API recommendations.

(2) Case 2 analysis. For some tasks, even when CAPIR cannot precisely decompose them into subtasks of optimal granularity, the process of task decomposition followed by API retrieval still enhances the overall effectiveness of API recommendation. For the task "Decompress and yield files with their paths from a given source DataPipe", the expected corresponding APIs for this task are ["IterDataPipe", "StreamWrapper", "Decompressor"].
Although CAPIR's decomposed subtasks may slightly deviate from the exact expectations, their more specific and diverse perspectives facilitate the retrieval and combination of APIs. This results in the correct APIs being ranked higher in the final recommendation list. Notably, the correct corresponding APIs, namely "Decompressor", "StreamWrapper", and "IterDataPipe" appear at the 2nd, 4th, and 6th positions of the CAPIR recommendation results.
In contrast, the ADA-retrieve results contain only "Decompressor" within the top-10 ranked APIs. This highlights the significance of CAPIR's task decomposition and retrieval approach in providing more accurate API recommendations, even in cases where precise task decomposition may not be fully achieved.

\subsection{CAPIR for Practical Application}

To validate the applicability of CAPIR for practical application, we engaged three graduate students with over five years of development experience, who had not seen the Torchdata-AR dataset before.
We provide them the Torchdata documentation to help them understand the functionality of APIs in this library.
Then, we require each of them to write 20 complex tasks (60 in total) that they think could be accomplished by Torchdata. We then filtered out 7 tasks that were not feasible to implement with Torchdata, leaving us with 53 tasks. Then, we wrote the corresponding APIs as recommendation targets. We compared  CAPIR with the two baselines on the Human-written datasets. The results are shown in Table \ref{tab:practical} (upper part). Compared with ADA-retrieve, CAPIR improves the \textit{Recall@10} by 75.5\% and \textit{Precision@10} by 75.6\%. Compared with CLEAR, CAPIR improves the \textit{Recall@10} by 69.3\% and \textit{Precision@10} by 70.6\%. The results demonstrate that, as long as the task descriptions are clear, CAPIR can accurately recommend the combination APIs for coarse-grained requirements.

Additionally, we selected 7 posts from the PyTorch community\footnote{https://discuss.pytorch.org/. Torchdata is developed by Pytorch community, so there are Q\&A posts about Torchdata in the Pytorch discussion board.} and StackOverflow, using the post titles or sentences describing requirements as tasks, and the corresponding APIs from the posts as recommendation targets. The experimental results are shown in Table \ref{tab:practical} (lower part).  Compared  with ADA-retrieve, CAPIR improves the \textit{Recall@10} from 35.7\% to 57.1\% and \textit{Precision@10} from 14.3\% to 22.9\%. Compared  with CLEAR, CAPIR improves 28.6\% for \textit{Recall@10} performance and 10.0\% for \textit{Precision@10}. These results demonstrate that CAPIR outperforms the baselines in recommending APIs for practical application.

\section{Related Work}

\subsection{API Recommendation}

API recommendation plays a crucial role in selecting suitable APIs that meet  user requirements. Two primary categories of API recommendation methods are information retrieval-based~\cite{huang2018api, raghothaman2016swim, rahman2016rack, wei2022clear, rahman2018nlp2api, liu2019generating, wei2023documentation, mcmillan2011portfolio,zhong2009mapo,buse2012synthesizing,gvero2015synthesizing} and learning-based~\cite{gu2016deep, martin2022deep, hadi2022effectiveness} approaches. 
Huang et al.~\cite{huang2018api} proposed BIKER, which retrieves the most similar SO posts by word embeddings model to bridge vocabulary gaps. Wei et al.~\cite{wei2022clear} proposed  CLEAR, which uses contrastive learning to train BERT~\cite{devlin2018bert} embedding model to retrieve APIs based on the similarity between SO posts and the query.  Some studies~\cite{zhong2009mapo, buse2012synthesizing, gvero2015synthesizing} recommend APIs by mining common API usage patterns. 
Gu et al.~\cite{gu2016deep} proposed DeepAPI, which used RNN~\cite{medsker2001recurrent} to generate relevant API sequence based on natural language queries. Martin et al.~\cite{martin2022deep} and Hadi et al.~\cite{hadi2022effectiveness} employ pre-trained models for end-to-end APIs generation.
Existing API recommendation methods heavily rely on large amounts of code and Q\&A forum data, and perform unsatisfactorily on libraries not seen in the training data. In contrast, our approach enables API recommendation  based only on  documentation.

\begin{table}
\centering

\caption{Evaluation results on API recommendation for practical application.\label{tab:practical}}
\resizebox{\linewidth}{!}{\begin{tabular}{cccccccc}
\hline
Dataset & Method & Recall@10 & Precision@10\\
\hline
\multirow{3}{*}{Human-written} 
&ADA-retrieve & 49.0\% & 17.2\% \\
&CLEAR & 50.8\% & 17.7\% \\
&\colorgray CAPIR (ours) & \colorgray \textbf{86.0\%} & \colorgray \textbf{30.2\%} \\
\hline
\multirow{3}{*}{Post} 
&ADA-retrieve & 35.7\% & 14.3\% \\
&CLEAR & 28.5\% & 12.9\% \\
&\colorgray CAPIR (ours) & \colorgray \textbf{57.1\%} & \colorgray \textbf{22.9\%} \\
\hline
\end{tabular}}
\end{table}

\subsection{Library-Oriented Code Generation}
Library-oriented code generation, which involves generating code
based on pre-defined APIs within a specific library, has recently
experienced significant advancements. 
Liu et al.~\cite{ase2023codegen4libs} proposed CodeGen4Libs, which incorporates two stages  to recommend the class-level APIs for a given requirement: First, they retrieve the most similar sample from the training set and extract the APIs it used. Then, the fine-tuned Generator module generates APIs corresponding to the requirement based on the retrieved APIs. CodeGen4Libs requires a large amount of training data as retrieval targets and to train the generator. This makes it inapplicable to libraries not present in the training data.
Zhou et al.~\cite{zhou2022docprompting} proposed Docprompting, which retrieves  relevant APIs from documents and adds them into prompts to improve code generation. Zan et al.~\cite{zan2022language} proposed CodeGenAPI, which retrieves private APIs from documents to improve code generation. Both Docprompting and CodeGenAPI directly retrieve APIs from documentation and cannot effectively handle coarse-grained tasks that require combinations of APIs. Moreover, both Docprompting and CodeGenAPI require a large amount of training data to train the retriever. Our baseline \textit{GPT-3.5 + ADA-retrieve} uses the same pipeline as they do, but with an off-the-shelf embedding, \textit{ada-embedding-002}.



\section{THREATS TO VALIDITY AND LIMITATIONS}

For our RAPID benchmark, we created developmental requirements using generated code summarizations. While these generated summarizations may exhibit some discrepancies compared to practical developmental requirements, we took  proactive steps to enhance the dataset's quality. Through manual annotation and careful filtering, we refined  the dataset to minimize the gap between the simulated scenarios and actual developmental requirements.



In our experiment, only Torchdata has low resources within the open-source community, while other libraries are widely used and extracted from existing datasets.This is mainly due to the significant human effort needed to build datasets for low-resource libraries. To ensure the fairness and validity of the experiment, we tested all libraries without any training.

We evaluated CAPIR on API recommendation for python and java language, demonstrating  that CAPIR can be applied across different languages. But we only evaluated CAPIR on libary-oriented code generation for python. As our method does not exploit Python-specific features, we believe that CAPIR could  be applied to library-oriented code generation in other programming languages as well.

\section{Conclusion}
In conclusion, we presented CAPIR, a novel compositional API recommendation approach for better aligning coarse-grained requirements with fine-grained APIs. CAPIR leverages the power of large language models to decompose coarse-grained tasks into fine-grained subtasks.
Then it retrieves the APIs corresponding to each subtask from the API documentation and uses a Reranker module to form the final recommendation results.
Moreover, we construct two benchmarks, RAPID and LOCG, to facilitate the evaluation of API recommendation methods on coarse-grained user requirements.
Through experiments on the RAPID benchmark, we demonstrated CAPIR outperforms baselines in API recommendation for coarse-grained requirements. The experiments on LOCG demonstrated that CAPIR can improve the library-oriented code generation performance.
With CAPIR, we envision a future that assistance tools can easily access relevant APIs for diverse software libraries and generate library-oriented code correctly for developers.

\begin{acks}
This work is supported by National Key Research and Development Program of China (Grant No. 2023YFB4503803).
\end{acks}

\bibliographystyle{ACM-Reference-Format}
\bibliography{reference}


\begin{thebibliography}{51}


\ifx \showCODEN    \undefined \def \showCODEN     #1{\unskip}     \fi
\ifx \showDOI      \undefined \def \showDOI       #1{#1}\fi
\ifx \showISBNx    \undefined \def \showISBNx     #1{\unskip}     \fi
\ifx \showISBNxiii \undefined \def \showISBNxiii  #1{\unskip}     \fi
\ifx \showISSN     \undefined \def \showISSN      #1{\unskip}     \fi
\ifx \showLCCN     \undefined \def \showLCCN      #1{\unskip}     \fi
\ifx \shownote     \undefined \def \shownote      #1{#1}          \fi
\ifx \showarticletitle \undefined \def \showarticletitle #1{#1}   \fi
\ifx \showURL      \undefined \def \showURL       {\relax}        \fi
\providecommand\bibfield[2]{#2}
\providecommand\bibinfo[2]{#2}
\providecommand\natexlab[1]{#1}
\providecommand\showeprint[2][]{arXiv:#2}

\bibitem[ADA(2022)]%
        {ADA-embedding-002}
 \bibinfo{year}{2022}\natexlab{}.
\newblock \bibinfo{title}{ADA Embedding}.
\newblock
  \bibinfo{howpublished}{\url{https://openai.com/blog/new-and-improved-embedding-model}}.
\newblock


\bibitem[cha(2022)]%
        {chatgpt}
 \bibinfo{year}{2022}\natexlab{}.
\newblock \bibinfo{title}{ChatGPT}.
\newblock \bibinfo{howpublished}{\url{https://chat.openai.com/}}.
\newblock


\bibitem[Cop(2022)]%
        {Copilot}
 \bibinfo{year}{2022}\natexlab{}.
\newblock \bibinfo{title}{GitHub Copilot}.
\newblock \bibinfo{howpublished}{\url{https://github.com/features/copilot}}.
\newblock


\bibitem[An et~al\mbox{.}(2023)]%
        {an2023incontext}
\bibfield{author}{\bibinfo{person}{Shengnan An}, \bibinfo{person}{Zeqi Lin},
  \bibinfo{person}{Qiang Fu}, \bibinfo{person}{Bei Chen},
  \bibinfo{person}{Nanning Zheng}, \bibinfo{person}{Jian-Guang Lou}, {and}
  \bibinfo{person}{Dongmei Zhang}.} \bibinfo{year}{2023}\natexlab{}.
\newblock \bibinfo{title}{How Do In-Context Examples Affect Compositional
  Generalization?}
\newblock
\newblock
\showeprint[arxiv]{2305.04835}~[cs.CL]


\bibitem[Barei{\ss} et~al\mbox{.}(2022)]%
        {bareiss2022code}
\bibfield{author}{\bibinfo{person}{Patrick Barei{\ss}},
  \bibinfo{person}{Beatriz Souza}, \bibinfo{person}{Marcelo d'Amorim}, {and}
  \bibinfo{person}{Michael Pradel}.} \bibinfo{year}{2022}\natexlab{}.
\newblock \showarticletitle{Code generation tools (almost) for free? a study of
  few-shot, pre-trained language models on code}.
\newblock \bibinfo{journal}{\emph{arXiv preprint arXiv:2206.01335}}
  (\bibinfo{year}{2022}).
\newblock


\bibitem[Barke et~al\mbox{.}(2023)]%
        {barke2023grounded}
\bibfield{author}{\bibinfo{person}{Shraddha Barke}, \bibinfo{person}{Michael~B
  James}, {and} \bibinfo{person}{Nadia Polikarpova}.}
  \bibinfo{year}{2023}\natexlab{}.
\newblock \showarticletitle{Grounded copilot: How programmers interact with
  code-generating models}.
\newblock \bibinfo{journal}{\emph{Proceedings of the ACM on Programming
  Languages}} \bibinfo{volume}{7}, \bibinfo{number}{OOPSLA1}
  (\bibinfo{year}{2023}), \bibinfo{pages}{85--111}.
\newblock


\bibitem[Brown et~al\mbox{.}(2020)]%
        {brown2020language}
\bibfield{author}{\bibinfo{person}{Tom Brown}, \bibinfo{person}{Benjamin Mann},
  \bibinfo{person}{Nick Ryder}, \bibinfo{person}{Melanie Subbiah},
  \bibinfo{person}{Jared~D Kaplan}, \bibinfo{person}{Prafulla Dhariwal},
  \bibinfo{person}{Arvind Neelakantan}, \bibinfo{person}{Pranav Shyam},
  \bibinfo{person}{Girish Sastry}, \bibinfo{person}{Amanda Askell},
  {et~al\mbox{.}}} \bibinfo{year}{2020}\natexlab{}.
\newblock \showarticletitle{Language models are few-shot learners}.
\newblock \bibinfo{journal}{\emph{Advances in neural information processing
  systems}}  \bibinfo{volume}{33} (\bibinfo{year}{2020}),
  \bibinfo{pages}{1877--1901}.
\newblock


\bibitem[Buse and Weimer(2012)]%
        {buse2012synthesizing}
\bibfield{author}{\bibinfo{person}{Raymond~PL Buse} {and}
  \bibinfo{person}{Westley Weimer}.} \bibinfo{year}{2012}\natexlab{}.
\newblock \showarticletitle{Synthesizing API usage examples}. In
  \bibinfo{booktitle}{\emph{2012 34th International Conference on Software
  Engineering (ICSE)}}. IEEE, \bibinfo{pages}{782--792}.
\newblock


\bibitem[Campello et~al\mbox{.}(2013)]%
        {campello2013density}
\bibfield{author}{\bibinfo{person}{Ricardo~JGB Campello},
  \bibinfo{person}{Davoud Moulavi}, {and} \bibinfo{person}{J{\"o}rg Sander}.}
  \bibinfo{year}{2013}\natexlab{}.
\newblock \showarticletitle{Density-based clustering based on hierarchical
  density estimates}. In \bibinfo{booktitle}{\emph{Pacific-Asia conference on
  knowledge discovery and data mining}}. Springer, \bibinfo{pages}{160--172}.
\newblock


\bibitem[Chan et~al\mbox{.}(2012)]%
        {chan2012searching}
\bibfield{author}{\bibinfo{person}{Wing-Kwan Chan}, \bibinfo{person}{Hong
  Cheng}, {and} \bibinfo{person}{David Lo}.} \bibinfo{year}{2012}\natexlab{}.
\newblock \showarticletitle{Searching connected API subgraph via text phrases}.
  In \bibinfo{booktitle}{\emph{Proceedings of the ACM SIGSOFT 20th
  International Symposium on the Foundations of Software Engineering}}.
  \bibinfo{pages}{1--11}.
\newblock


\bibitem[Chen et~al\mbox{.}(2022)]%
        {chen2022codet}
\bibfield{author}{\bibinfo{person}{Bei Chen}, \bibinfo{person}{Fengji Zhang},
  \bibinfo{person}{Anh Nguyen}, \bibinfo{person}{Daoguang Zan},
  \bibinfo{person}{Zeqi Lin}, \bibinfo{person}{Jian-Guang Lou}, {and}
  \bibinfo{person}{Weizhu Chen}.} \bibinfo{year}{2022}\natexlab{}.
\newblock \showarticletitle{Codet: Code generation with generated tests}.
\newblock \bibinfo{journal}{\emph{arXiv preprint arXiv:2207.10397}}
  (\bibinfo{year}{2022}).
\newblock


\bibitem[Chen et~al\mbox{.}(2021)]%
        {chen2021evaluating}
\bibfield{author}{\bibinfo{person}{Mark Chen}, \bibinfo{person}{Jerry Tworek},
  \bibinfo{person}{Heewoo Jun}, \bibinfo{person}{Qiming Yuan},
  \bibinfo{person}{Henrique Ponde de~Oliveira Pinto}, \bibinfo{person}{Jared
  Kaplan}, \bibinfo{person}{Harri Edwards}, \bibinfo{person}{Yuri Burda},
  \bibinfo{person}{Nicholas Joseph}, \bibinfo{person}{Greg Brockman},
  {et~al\mbox{.}}} \bibinfo{year}{2021}\natexlab{}.
\newblock \showarticletitle{Evaluating large language models trained on code}.
\newblock \bibinfo{journal}{\emph{arXiv preprint arXiv:2107.03374}}
  (\bibinfo{year}{2021}).
\newblock


\bibitem[Ciniselli et~al\mbox{.}(2023)]%
        {ciniselli2023source}
\bibfield{author}{\bibinfo{person}{Matteo Ciniselli}, \bibinfo{person}{Luca
  Pascarella}, \bibinfo{person}{Emad Aghajani}, \bibinfo{person}{Simone
  Scalabrino}, \bibinfo{person}{Rocco Oliveto}, {and} \bibinfo{person}{Gabriele
  Bavota}.} \bibinfo{year}{2023}\natexlab{}.
\newblock \showarticletitle{Source Code Recommender Systems: The Practitioners'
  Perspective}.
\newblock \bibinfo{journal}{\emph{arXiv preprint arXiv:2302.04098}}
  (\bibinfo{year}{2023}).
\newblock


\bibitem[Devlin et~al\mbox{.}(2018)]%
        {devlin2018bert}
\bibfield{author}{\bibinfo{person}{Jacob Devlin}, \bibinfo{person}{Ming-Wei
  Chang}, \bibinfo{person}{Kenton Lee}, {and} \bibinfo{person}{Kristina
  Toutanova}.} \bibinfo{year}{2018}\natexlab{}.
\newblock \showarticletitle{Bert: Pre-training of deep bidirectional
  transformers for language understanding}.
\newblock \bibinfo{journal}{\emph{arXiv preprint arXiv:1810.04805}}
  (\bibinfo{year}{2018}).
\newblock


\bibitem[Dong et~al\mbox{.}(2022)]%
        {dong2022survey}
\bibfield{author}{\bibinfo{person}{Qingxiu Dong}, \bibinfo{person}{Lei Li},
  \bibinfo{person}{Damai Dai}, \bibinfo{person}{Ce Zheng},
  \bibinfo{person}{Zhiyong Wu}, \bibinfo{person}{Baobao Chang},
  \bibinfo{person}{Xu Sun}, \bibinfo{person}{Jingjing Xu}, {and}
  \bibinfo{person}{Zhifang Sui}.} \bibinfo{year}{2022}\natexlab{}.
\newblock \showarticletitle{A survey for in-context learning}.
\newblock \bibinfo{journal}{\emph{arXiv preprint arXiv:2301.00234}}
  (\bibinfo{year}{2022}).
\newblock


\bibitem[Floridi and Chiriatti(2020)]%
        {floridi2020gpt}
\bibfield{author}{\bibinfo{person}{Luciano Floridi} {and}
  \bibinfo{person}{Massimo Chiriatti}.} \bibinfo{year}{2020}\natexlab{}.
\newblock \showarticletitle{GPT-3: Its nature, scope, limits, and
  consequences}.
\newblock \bibinfo{journal}{\emph{Minds and Machines}}  \bibinfo{volume}{30}
  (\bibinfo{year}{2020}), \bibinfo{pages}{681--694}.
\newblock


\bibitem[Gu et~al\mbox{.}(2016)]%
        {gu2016deep}
\bibfield{author}{\bibinfo{person}{Xiaodong Gu}, \bibinfo{person}{Hongyu
  Zhang}, \bibinfo{person}{Dongmei Zhang}, {and} \bibinfo{person}{Sunghun
  Kim}.} \bibinfo{year}{2016}\natexlab{}.
\newblock \showarticletitle{Deep API learning}. In
  \bibinfo{booktitle}{\emph{Proceedings of the 2016 24th ACM SIGSOFT
  international symposium on foundations of software engineering}}.
  \bibinfo{pages}{631--642}.
\newblock


\bibitem[Gvero and Kuncak(2015)]%
        {gvero2015synthesizing}
\bibfield{author}{\bibinfo{person}{Tihomir Gvero} {and} \bibinfo{person}{Viktor
  Kuncak}.} \bibinfo{year}{2015}\natexlab{}.
\newblock \showarticletitle{Synthesizing Java expressions from free-form
  queries}. In \bibinfo{booktitle}{\emph{Proceedings of the 2015 acm sigplan
  international conference on object-oriented programming, systems, languages,
  and applications}}. \bibinfo{pages}{416--432}.
\newblock


\bibitem[Hadi et~al\mbox{.}(2022)]%
        {hadi2022effectiveness}
\bibfield{author}{\bibinfo{person}{Mohammad~Abdul Hadi}, \bibinfo{person}{Imam
  Nur~Bani Yusuf}, \bibinfo{person}{Ferdian Thung}, \bibinfo{person}{Kien~Gia
  Luong}, \bibinfo{person}{Jiang Lingxiao}, \bibinfo{person}{Fatemeh~H Fard},
  {and} \bibinfo{person}{David Lo}.} \bibinfo{year}{2022}\natexlab{}.
\newblock \showarticletitle{On the effectiveness of pretrained models for api
  learning}. In \bibinfo{booktitle}{\emph{Proceedings of the 30th IEEE/ACM
  International Conference on Program Comprehension}}.
  \bibinfo{pages}{309--320}.
\newblock


\bibitem[He et~al\mbox{.}(2021)]%
        {he2021pyart}
\bibfield{author}{\bibinfo{person}{Xincheng He}, \bibinfo{person}{Lei Xu},
  \bibinfo{person}{Xiangyu Zhang}, \bibinfo{person}{Rui Hao},
  \bibinfo{person}{Yang Feng}, {and} \bibinfo{person}{Baowen Xu}.}
  \bibinfo{year}{2021}\natexlab{}.
\newblock \showarticletitle{Pyart: Python api recommendation in real-time}. In
  \bibinfo{booktitle}{\emph{2021 IEEE/ACM 43rd International Conference on
  Software Engineering (ICSE)}}. IEEE, \bibinfo{pages}{1634--1645}.
\newblock


\bibitem[Huang et~al\mbox{.}(2018)]%
        {huang2018api}
\bibfield{author}{\bibinfo{person}{Qiao Huang}, \bibinfo{person}{Xin Xia},
  \bibinfo{person}{Zhenchang Xing}, \bibinfo{person}{David Lo}, {and}
  \bibinfo{person}{Xinyu Wang}.} \bibinfo{year}{2018}\natexlab{}.
\newblock \showarticletitle{API method recommendation without worrying about
  the task-API knowledge gap}. In \bibinfo{booktitle}{\emph{Proceedings of the
  33rd ACM/IEEE International Conference on Automated Software Engineering}}.
  \bibinfo{pages}{293--304}.
\newblock


\bibitem[Irsan et~al\mbox{.}(2023)]%
        {irsan2023multi}
\bibfield{author}{\bibinfo{person}{Ivana~Clairine Irsan}, \bibinfo{person}{Ting
  Zhang}, \bibinfo{person}{Ferdian Thung}, \bibinfo{person}{Kisub Kim}, {and}
  \bibinfo{person}{David Lo}.} \bibinfo{year}{2023}\natexlab{}.
\newblock \showarticletitle{Multi-Modal API Recommendation}. In
  \bibinfo{booktitle}{\emph{2023 IEEE International Conference on Software
  Analysis, Evolution and Reengineering (SANER)}}. IEEE,
  \bibinfo{pages}{272--283}.
\newblock


\bibitem[Kojima et~al\mbox{.}(2022)]%
        {kojima2022large}
\bibfield{author}{\bibinfo{person}{Takeshi Kojima},
  \bibinfo{person}{Shixiang~Shane Gu}, \bibinfo{person}{Machel Reid},
  \bibinfo{person}{Yutaka Matsuo}, {and} \bibinfo{person}{Yusuke Iwasawa}.}
  \bibinfo{year}{2022}\natexlab{}.
\newblock \showarticletitle{Large language models are zero-shot reasoners}.
\newblock \bibinfo{journal}{\emph{Advances in neural information processing
  systems}}  \bibinfo{volume}{35} (\bibinfo{year}{2022}),
  \bibinfo{pages}{22199--22213}.
\newblock


\bibitem[Li et~al\mbox{.}(2022)]%
        {li2022competition}
\bibfield{author}{\bibinfo{person}{Yujia Li}, \bibinfo{person}{David Choi},
  \bibinfo{person}{Junyoung Chung}, \bibinfo{person}{Nate Kushman},
  \bibinfo{person}{Julian Schrittwieser}, \bibinfo{person}{R{\'e}mi Leblond},
  \bibinfo{person}{Tom Eccles}, \bibinfo{person}{James Keeling},
  \bibinfo{person}{Felix Gimeno}, \bibinfo{person}{Agustin Dal~Lago},
  {et~al\mbox{.}}} \bibinfo{year}{2022}\natexlab{}.
\newblock \showarticletitle{Competition-level code generation with alphacode}.
\newblock \bibinfo{journal}{\emph{Science}} \bibinfo{volume}{378},
  \bibinfo{number}{6624} (\bibinfo{year}{2022}), \bibinfo{pages}{1092--1097}.
\newblock


\bibitem[Li et~al\mbox{.}(2023)]%
        {li2023making}
\bibfield{author}{\bibinfo{person}{Yifei Li}, \bibinfo{person}{Zeqi Lin},
  \bibinfo{person}{Shizhuo Zhang}, \bibinfo{person}{Qiang Fu},
  \bibinfo{person}{Bei Chen}, \bibinfo{person}{Jian-Guang Lou}, {and}
  \bibinfo{person}{Weizhu Chen}.} \bibinfo{year}{2023}\natexlab{}.
\newblock \showarticletitle{Making language models better reasoners with
  step-aware verifier}. In \bibinfo{booktitle}{\emph{Proceedings of the 61st
  Annual Meeting of the Association for Computational Linguistics (Volume 1:
  Long Papers)}}. \bibinfo{pages}{5315--5333}.
\newblock


\bibitem[Liu et~al\mbox{.}(2019)]%
        {liu2019generating}
\bibfield{author}{\bibinfo{person}{Mingwei Liu}, \bibinfo{person}{Xin Peng},
  \bibinfo{person}{Andrian Marcus}, \bibinfo{person}{Zhenchang Xing},
  \bibinfo{person}{Wenkai Xie}, \bibinfo{person}{Shuangshuang Xing}, {and}
  \bibinfo{person}{Yang Liu}.} \bibinfo{year}{2019}\natexlab{}.
\newblock \showarticletitle{Generating query-specific class API summaries}. In
  \bibinfo{booktitle}{\emph{Proceedings of the 2019 27th ACM joint meeting on
  European software engineering conference and symposium on the foundations of
  software engineering}}. \bibinfo{pages}{120--130}.
\newblock


\bibitem[Liu et~al\mbox{.}(2023)]%
        {ase2023codegen4libs}
\bibfield{author}{\bibinfo{person}{Mingwei Liu}, \bibinfo{person}{Tianyong
  Yang}, \bibinfo{person}{Yiling Lou}, \bibinfo{person}{Xueying Du},
  \bibinfo{person}{Ying Wang}, \bibinfo{person}{}, {and} \bibinfo{person}{Xin
  Peng}.} \bibinfo{year}{2023}\natexlab{}.
\newblock \showarticletitle{{CodeGen4Libs}: A Two-stage Approach for
  Library-oriented Code Generation}. In \bibinfo{booktitle}{\emph{38th
  {IEEE/ACM} International Conference on Automated Software Engineering, {ASE}
  2023, Kirchberg, Luxembourg, September 11-15, 2023}}.
  \bibinfo{publisher}{{IEEE}}, \bibinfo{pages}{0--0}.
\newblock


\bibitem[Martin and Guo(2022)]%
        {martin2022deep}
\bibfield{author}{\bibinfo{person}{James Martin} {and} \bibinfo{person}{Jin~LC
  Guo}.} \bibinfo{year}{2022}\natexlab{}.
\newblock \showarticletitle{Deep api learning revisited}. In
  \bibinfo{booktitle}{\emph{Proceedings of the 30th IEEE/ACM International
  Conference on Program Comprehension}}. \bibinfo{pages}{321--330}.
\newblock


\bibitem[McInnes et~al\mbox{.}(2018)]%
        {mcinnes2018umap}
\bibfield{author}{\bibinfo{person}{Leland McInnes}, \bibinfo{person}{John
  Healy}, {and} \bibinfo{person}{James Melville}.}
  \bibinfo{year}{2018}\natexlab{}.
\newblock \showarticletitle{Umap: Uniform manifold approximation and projection
  for dimension reduction}.
\newblock \bibinfo{journal}{\emph{arXiv preprint arXiv:1802.03426}}
  (\bibinfo{year}{2018}).
\newblock


\bibitem[McMillan et~al\mbox{.}(2011)]%
        {mcmillan2011portfolio}
\bibfield{author}{\bibinfo{person}{Collin McMillan}, \bibinfo{person}{Mark
  Grechanik}, \bibinfo{person}{Denys Poshyvanyk}, \bibinfo{person}{Qing Xie},
  {and} \bibinfo{person}{Chen Fu}.} \bibinfo{year}{2011}\natexlab{}.
\newblock \showarticletitle{Portfolio: finding relevant functions and their
  usage}. In \bibinfo{booktitle}{\emph{Proceedings of the 33rd International
  Conference on Software Engineering}}. \bibinfo{pages}{111--120}.
\newblock


\bibitem[Medsker and Jain(2001)]%
        {medsker2001recurrent}
\bibfield{author}{\bibinfo{person}{Larry~R Medsker} {and} \bibinfo{person}{LC
  Jain}.} \bibinfo{year}{2001}\natexlab{}.
\newblock \showarticletitle{Recurrent neural networks}.
\newblock \bibinfo{journal}{\emph{Design and Applications}}
  \bibinfo{volume}{5}, \bibinfo{number}{64-67} (\bibinfo{year}{2001}),
  \bibinfo{pages}{2}.
\newblock


\bibitem[OpenAI(2023)]%
        {openai2023gpt4}
\bibfield{author}{\bibinfo{person}{OpenAI}.} \bibinfo{year}{2023}\natexlab{}.
\newblock \bibinfo{title}{GPT-4 Technical Report}.
\newblock
\newblock
\showeprint[arxiv]{2303.08774}~[cs.CL]


\bibitem[Peng et~al\mbox{.}(2022)]%
        {peng2022revisiting}
\bibfield{author}{\bibinfo{person}{Yun Peng}, \bibinfo{person}{Shuqing Li},
  \bibinfo{person}{Wenwei Gu}, \bibinfo{person}{Yichen Li},
  \bibinfo{person}{Wenxuan Wang}, \bibinfo{person}{Cuiyun Gao}, {and}
  \bibinfo{person}{Michael~R Lyu}.} \bibinfo{year}{2022}\natexlab{}.
\newblock \showarticletitle{Revisiting, benchmarking and exploring API
  recommendation: How far are we?}
\newblock \bibinfo{journal}{\emph{IEEE Transactions on Software Engineering}}
  \bibinfo{volume}{49}, \bibinfo{number}{4} (\bibinfo{year}{2022}),
  \bibinfo{pages}{1876--1897}.
\newblock


\bibitem[Raghothaman et~al\mbox{.}(2016)]%
        {raghothaman2016swim}
\bibfield{author}{\bibinfo{person}{Mukund Raghothaman}, \bibinfo{person}{Yi
  Wei}, {and} \bibinfo{person}{Youssef Hamadi}.}
  \bibinfo{year}{2016}\natexlab{}.
\newblock \showarticletitle{SWIM: synthesizing what I mean: code search and
  idiomatic snippet synthesis}. In \bibinfo{booktitle}{\emph{Proceedings of the
  38th International Conference on Software Engineering}}.
  \bibinfo{pages}{357--367}.
\newblock


\bibitem[Rahman and Roy(2018)]%
        {rahman2018nlp2api}
\bibfield{author}{\bibinfo{person}{Mohammad~Masudur Rahman} {and}
  \bibinfo{person}{Chanchal Roy}.} \bibinfo{year}{2018}\natexlab{}.
\newblock \showarticletitle{Nlp2api: Query reformulation for code search using
  crowdsourced knowledge and extra-large data analytics}. In
  \bibinfo{booktitle}{\emph{2018 IEEE International Conference on Software
  Maintenance and Evolution (ICSME)}}. IEEE, \bibinfo{pages}{714--714}.
\newblock


\bibitem[Rahman et~al\mbox{.}(2016)]%
        {rahman2016rack}
\bibfield{author}{\bibinfo{person}{Mohammad~Masudur Rahman},
  \bibinfo{person}{Chanchal~K Roy}, {and} \bibinfo{person}{David Lo}.}
  \bibinfo{year}{2016}\natexlab{}.
\newblock \showarticletitle{Rack: Automatic api recommendation using
  crowdsourced knowledge}. In \bibinfo{booktitle}{\emph{2016 IEEE 23rd
  International Conference on Software Analysis, Evolution, and Reengineering
  (SANER)}}, Vol.~\bibinfo{volume}{1}. IEEE, \bibinfo{pages}{349--359}.
\newblock


\bibitem[Sun et~al\mbox{.}(2023)]%
        {sun2023automatic}
\bibfield{author}{\bibinfo{person}{Weisong Sun}, \bibinfo{person}{Chunrong
  Fang}, \bibinfo{person}{Yudu You}, \bibinfo{person}{Yun Miao},
  \bibinfo{person}{Yi Liu}, \bibinfo{person}{Yuekang Li},
  \bibinfo{person}{Gelei Deng}, \bibinfo{person}{Shenghan Huang},
  \bibinfo{person}{Yuchen Chen}, \bibinfo{person}{Quanjun Zhang},
  {et~al\mbox{.}}} \bibinfo{year}{2023}\natexlab{}.
\newblock \showarticletitle{Automatic Code Summarization via ChatGPT: How Far
  Are We?}
\newblock \bibinfo{journal}{\emph{arXiv preprint arXiv:2305.12865}}
  (\bibinfo{year}{2023}).
\newblock


\bibitem[Vaithilingam et~al\mbox{.}(2022)]%
        {vaithilingam2022expectation}
\bibfield{author}{\bibinfo{person}{Priyan Vaithilingam},
  \bibinfo{person}{Tianyi Zhang}, {and} \bibinfo{person}{Elena~L Glassman}.}
  \bibinfo{year}{2022}\natexlab{}.
\newblock \showarticletitle{Expectation vs. experience: Evaluating the
  usability of code generation tools powered by large language models}. In
  \bibinfo{booktitle}{\emph{Chi conference on human factors in computing
  systems extended abstracts}}. \bibinfo{pages}{1--7}.
\newblock


\bibitem[Wang et~al\mbox{.}(2022a)]%
        {wang2022self}
\bibfield{author}{\bibinfo{person}{Xuezhi Wang}, \bibinfo{person}{Jason Wei},
  \bibinfo{person}{Dale Schuurmans}, \bibinfo{person}{Quoc~V Le},
  \bibinfo{person}{Ed~H Chi}, \bibinfo{person}{Sharan Narang},
  \bibinfo{person}{Aakanksha Chowdhery}, {and} \bibinfo{person}{Denny Zhou}.}
  \bibinfo{year}{2022}\natexlab{a}.
\newblock \showarticletitle{Self-Consistency Improves Chain of Thought
  Reasoning in Language Models}. In \bibinfo{booktitle}{\emph{The Eleventh
  International Conference on Learning Representations}}.
\newblock


\bibitem[Wang et~al\mbox{.}(2022b)]%
        {wang2022execution}
\bibfield{author}{\bibinfo{person}{Zhiruo Wang}, \bibinfo{person}{Shuyan Zhou},
  \bibinfo{person}{Daniel Fried}, {and} \bibinfo{person}{Graham Neubig}.}
  \bibinfo{year}{2022}\natexlab{b}.
\newblock \showarticletitle{Execution-based evaluation for open-domain code
  generation}.
\newblock \bibinfo{journal}{\emph{arXiv preprint arXiv:2212.10481}}
  (\bibinfo{year}{2022}).
\newblock


\bibitem[Wei et~al\mbox{.}(2023)]%
        {wei2023documentation}
\bibfield{author}{\bibinfo{person}{Hongwei Wei}, \bibinfo{person}{Xiaohong Su},
  \bibinfo{person}{Weining Zheng}, {and} \bibinfo{person}{Wenxin Tao}.}
  \bibinfo{year}{2023}\natexlab{}.
\newblock \showarticletitle{Documentation-Guided API Sequence Search without
  Worrying about the Text-API Semantic Gap}. In \bibinfo{booktitle}{\emph{2023
  IEEE International Conference on Software Analysis, Evolution and
  Reengineering (SANER)}}. IEEE, \bibinfo{pages}{343--354}.
\newblock


\bibitem[Wei et~al\mbox{.}(2022c)]%
        {wei2022chain}
\bibfield{author}{\bibinfo{person}{Jason Wei}, \bibinfo{person}{Xuezhi Wang},
  \bibinfo{person}{Dale Schuurmans}, \bibinfo{person}{Maarten Bosma},
  \bibinfo{person}{Fei Xia}, \bibinfo{person}{Ed Chi}, \bibinfo{person}{Quoc~V
  Le}, \bibinfo{person}{Denny Zhou}, {et~al\mbox{.}}}
  \bibinfo{year}{2022}\natexlab{c}.
\newblock \showarticletitle{Chain-of-thought prompting elicits reasoning in
  large language models}.
\newblock \bibinfo{journal}{\emph{Advances in Neural Information Processing
  Systems}}  \bibinfo{volume}{35} (\bibinfo{year}{2022}),
  \bibinfo{pages}{24824--24837}.
\newblock


\bibitem[Wei et~al\mbox{.}(2022a)]%
        {wei2022clear}
\bibfield{author}{\bibinfo{person}{Moshi Wei}, \bibinfo{person}{Nima~Shiri
  Harzevili}, \bibinfo{person}{Yuchao Huang}, \bibinfo{person}{Junjie Wang},
  {and} \bibinfo{person}{Song Wang}.} \bibinfo{year}{2022}\natexlab{a}.
\newblock \showarticletitle{Clear: contrastive learning for api
  recommendation}. In \bibinfo{booktitle}{\emph{Proceedings of the 44th
  International Conference on Software Engineering}}.
  \bibinfo{pages}{376--387}.
\newblock


\bibitem[Wei et~al\mbox{.}(2022b)]%
        {wei2022api}
\bibfield{author}{\bibinfo{person}{Moshi Wei}, \bibinfo{person}{Yuchao Huang},
  \bibinfo{person}{Junjie Wang}, \bibinfo{person}{Jiho Shin},
  \bibinfo{person}{Nima~Shiri Harzevili}, {and} \bibinfo{person}{Song Wang}.}
  \bibinfo{year}{2022}\natexlab{b}.
\newblock \showarticletitle{API recommendation for machine learning libraries:
  how far are we?}. In \bibinfo{booktitle}{\emph{Proceedings of the 30th ACM
  Joint European Software Engineering Conference and Symposium on the
  Foundations of Software Engineering}}. \bibinfo{pages}{370--381}.
\newblock


\bibitem[Yin et~al\mbox{.}(2018)]%
        {yin2018learning}
\bibfield{author}{\bibinfo{person}{Pengcheng Yin}, \bibinfo{person}{Bowen
  Deng}, \bibinfo{person}{Edgar Chen}, \bibinfo{person}{Bogdan Vasilescu},
  {and} \bibinfo{person}{Graham Neubig}.} \bibinfo{year}{2018}\natexlab{}.
\newblock \showarticletitle{Learning to mine aligned code and natural language
  pairs from stack overflow}. In \bibinfo{booktitle}{\emph{Proceedings of the
  15th international conference on mining software repositories}}.
  \bibinfo{pages}{476--486}.
\newblock


\bibitem[Zan et~al\mbox{.}(2023)]%
        {zan2023private}
\bibfield{author}{\bibinfo{person}{Daoguang Zan}, \bibinfo{person}{Bei Chen},
  \bibinfo{person}{Yongshun Gong}, \bibinfo{person}{Junzhi Cao},
  \bibinfo{person}{Fengji Zhang}, \bibinfo{person}{Bingchao Wu},
  \bibinfo{person}{Bei Guan}, \bibinfo{person}{Yilong Yin}, {and}
  \bibinfo{person}{Yongji Wang}.} \bibinfo{year}{2023}\natexlab{}.
\newblock \showarticletitle{Private-library-oriented code generation with large
  language models}.
\newblock \bibinfo{journal}{\emph{arXiv preprint arXiv:2307.15370}}
  (\bibinfo{year}{2023}).
\newblock


\bibitem[Zan et~al\mbox{.}(2022a)]%
        {zan2022language}
\bibfield{author}{\bibinfo{person}{Daoguang Zan}, \bibinfo{person}{Bei Chen},
  \bibinfo{person}{Zeqi Lin}, \bibinfo{person}{Bei Guan},
  \bibinfo{person}{Yongji Wang}, {and} \bibinfo{person}{Jian-Guang Lou}.}
  \bibinfo{year}{2022}\natexlab{a}.
\newblock \showarticletitle{When language model meets private library}.
\newblock \bibinfo{journal}{\emph{arXiv preprint arXiv:2210.17236}}
  (\bibinfo{year}{2022}).
\newblock


\bibitem[Zan et~al\mbox{.}(2022b)]%
        {zan2022cert}
\bibfield{author}{\bibinfo{person}{Daoguang Zan}, \bibinfo{person}{Bei Chen},
  \bibinfo{person}{Dejian Yang}, \bibinfo{person}{Zeqi Lin},
  \bibinfo{person}{Minsu Kim}, \bibinfo{person}{Bei Guan},
  \bibinfo{person}{Yongji Wang}, \bibinfo{person}{Weizhu Chen}, {and}
  \bibinfo{person}{Jian-Guang Lou}.} \bibinfo{year}{2022}\natexlab{b}.
\newblock \showarticletitle{CERT: Continual Pre-training on Sketches for
  Library-oriented Code Generation}.
\newblock \bibinfo{journal}{\emph{arXiv preprint arXiv:2206.06888}}
  (\bibinfo{year}{2022}).
\newblock


\bibitem[Zhang et~al\mbox{.}(2023)]%
        {zhang2023coder}
\bibfield{author}{\bibinfo{person}{Tianyi Zhang}, \bibinfo{person}{Tao Yu},
  \bibinfo{person}{Tatsunori Hashimoto}, \bibinfo{person}{Mike Lewis},
  \bibinfo{person}{Wen-tau Yih}, \bibinfo{person}{Daniel Fried}, {and}
  \bibinfo{person}{Sida Wang}.} \bibinfo{year}{2023}\natexlab{}.
\newblock \showarticletitle{Coder reviewer reranking for code generation}. In
  \bibinfo{booktitle}{\emph{International Conference on Machine Learning}}.
  PMLR, \bibinfo{pages}{41832--41846}.
\newblock


\bibitem[Zhong et~al\mbox{.}(2009)]%
        {zhong2009mapo}
\bibfield{author}{\bibinfo{person}{Hao Zhong}, \bibinfo{person}{Tao Xie},
  \bibinfo{person}{Lu Zhang}, \bibinfo{person}{Jian Pei}, {and}
  \bibinfo{person}{Hong Mei}.} \bibinfo{year}{2009}\natexlab{}.
\newblock \showarticletitle{MAPO: Mining and recommending API usage patterns}.
  In \bibinfo{booktitle}{\emph{ECOOP 2009--Object-Oriented Programming: 23rd
  European Conference, Genoa, Italy, July 6-10, 2009. Proceedings 23}}.
  Springer, \bibinfo{pages}{318--343}.
\newblock


\bibitem[Zhou et~al\mbox{.}(2022)]%
        {zhou2022docprompting}
\bibfield{author}{\bibinfo{person}{Shuyan Zhou}, \bibinfo{person}{Uri Alon},
  \bibinfo{person}{Frank~F Xu}, \bibinfo{person}{Zhengbao Jiang}, {and}
  \bibinfo{person}{Graham Neubig}.} \bibinfo{year}{2022}\natexlab{}.
\newblock \showarticletitle{Docprompting: Generating code by retrieving the
  docs}. In \bibinfo{booktitle}{\emph{The Eleventh International Conference on
  Learning Representations}}.
\newblock


\end{thebibliography}

\end{document}